\documentclass[acmsmall,nonacm,10pt]{acmart}
\pdfoutput=1
\renewcommand\footnotetextcopyrightpermission[1]{}

\acmJournal{PACMPL}

\setcopyright{none}

\usepackage{subcaption} 

\usepackage{booktabs}

\usepackage{xcolor}

\usepackage{flushend}

\usepackage{relsize}
\usepackage{paralist}

\usepackage{adjustbox}

\usepackage{url}
\usepackage{epigraph}
\usepackage{pdfpages}
\usepackage{wrapfig}
\usepackage{multicol}
\usepackage{multirow}
\usepackage[flushleft]{threeparttable}

\usepackage{mathtools}

\usepackage{mathpartir}

\usepackage[tableposition = top]{caption}

\usepackage{amsmath}

\newcommand{\twopartdef}[3] {
\left\{\!\!\!
\begin{array}{ll}
   #1 & \mbox{if \ } #2 \\
   #3 & \mbox{else}
\end{array}
\right.
}

\usepackage{amssymb}

\usepackage{subcaption}

\hypersetup{
    linktoc=all,
    colorlinks=false,       
    linkcolor=black,          
    linkbordercolor=black,
    citecolor=black,        
    filecolor=black,      
    urlcolor=black,           
    pdfborderstyle={/S/U/W 0}
}

\usepackage{tikz}

\usepackage{array}
\usepackage{listings}
\usepackage{microtype}
\usepackage{bm}
\usepackage{enumitem}

\usepackage{mmm}
\usepackage{pl}
\usepackage{mathpartir}

\usepackage{amsthm}

\newtheorem{definition}{Definition}
\newtheorem{theorem}{Theorem}
\newtheorem{lemma}{Lemma}

\newtheorem*{theorem*}{Theorem}

\newcolumntype{?}{!{\color{white}\vrule width 1pt}}
\usepackage[utf8]{inputenc}
\usepackage{textcomp}
\usepackage{colortbl}
\usepackage{pgfplots}
\usepackage{pgfplotstable}

\let\OldS\S

\newcommand{\co}[2]{#2}

\usepackage{mathtools}

\usepackage{tabularx}

\def\t{\phantom{XX}}
\def\st{\phantom{X}}

\newcommand{\union}{\cup}

\newcommand{\intersect}{\cap}

\def\product{\cdot}
\def\s.t.{\;|\;}

\def\<{\langle}
\def\>{\rangle}

\def\free{\mathsf{free}}

\def\pushi{\mathsf{push+}}
\def\pulli{\mathsf{pull+}}
\def\pushn{\mathsf{push-}}
\def\pulln{\mathsf{pull-}}

\def\simple{\mathsf{simple}}

\def\D{\mathcal{D}}
\def\R{\mathcal{R}}
\def\F{\mathcal{F}}
\def\G{\mathcal{G}}
\def\E{\mathcal{E}}
\def\P{\mathcal{P}}
\def\C{\mathcal{C}}
\def\CB{\mathbb{C}}
\def\I{\mathcal{I}}
\def\S{\mathcal{S}}
\def\B{\mathcal{B}}

\def\CR{\mathbb{R}}
\def\CM{\mathbb{M}}
\def\CRs{\mathbb{R}\mathsf{s}}
\def\CMs{\mathbb{M}\mathsf{s}}

\def\V{\mathsf{V}}

\usepackage{centernot}

\def\lif{\mathsf{if }}
\def\lthen{\mathsf{\ then \ }}
\def\lelse{\mathsf{else }}

\def\llet{\mathsf{let \ }}
\def\ilet{\mathsf{ilet \ }}
\def\mlet{\mathsf{mlet \ }}
\def\rlet{\mathsf{rlet \ }}

\def\lin{\mathsf{\ in \ }}

\def\tail{\mathsf{tail}}
\def\lreturn{\mathsf{return }}

\def\True{\mathsf{True}}
\def\False{\mathsf{False}}

\def\Nat{\mathbb{N}}

\def\head{\mathsf{head}}

\def\tail{\mathsf{tail}}

\def\Spec{\mathcal{S}pec}

\def\Radius{\textsc{Radius}}

\def\NWR{\textsc{NWR}}
\def\Trust{\textsc{Trust}}

\def\WSP{\textsc{WSP}}

\def\SP{\textsc{SSSP}}

\def\SSSP{\textsc{SSSP}}

\def\WP{\textsc{WP}}
\def\BFS{\textsc{BFS}}
\def\CC{\textsc{CC}}

\def\WP{\textsc{WP}}

\def\NSP{\textsc{NSP}}

\def\DRR{\textsc{DRR}}
\def\DS{\textsc{DS}}
\def\RDS{\textsc{RDS}}

\def\Vertices{\mathsf{V}}
\def\PathSet{\mathsf{P}}
\def\Paths{\mathsf{Paths}}

\def\n{\mathsf{n}}
\def\v{\mathsf{v}}

\DeclareMathOperator*{\argmin}{arg\,min}
\DeclareMathOperator*{\argsmin}{args\,min}

\def\head{\mathsf{head}}
\def\tail{\mathsf{tail}}
\def\weight{\mathsf{weight}}
\def\capacity{\mathsf{capacity}}
\def\length{\mathsf{length}}

\def\path{\mathsf{path}}

\def\penultimate{\mathsf{penultimate}}

\def\preds{\mathsf{preds}}
\def\CPreds{\mathsf{CPreds}}

\newcommand*{\fullref}[1]{\hyperref[{#1}]{\autoref*{#1} (\nameref*{#1})}}

\AtEndDocument{\refstepcounter{definition}\label{finaldefinition}}
\AtEndDocument{\refstepcounter{theorem}\label{finaltheorem}}
\AtEndDocument{\refstepcounter{llemma}\label{finallemma}}

\newcommand{\undersetp}[2]{{#2}_{#1}}

\usepackage{dashbox}

\usepackage{stmaryrd}
\newcommand{\sem}[1]{\left\llbracket \, #1 \, \right\rrbracket}

\makeatletter
\newcommand{\rulelabel}[1]{%
   \protected@write \@auxout {}{\string \newlabel {#1}{{\textsc{#1}}{\thepage}{}{#1}{}} }%
   \hypertarget{#1}{}
}
\makeatother

\newcommand{\irule}[1]{\ref{#1}}

\makeatletter
\newcommand{\rulelabelp}[2]{%
   \protected@write \@auxout {}{\string \newlabel {#1}{{\textsc{#2}}{\thepage}{#2}{#1}{}} }%
   \hypertarget{#1}{}
}
\makeatother

\usepackage{dashbox}

\def\name{\textsc{Grafs}}

\definecolor{ForestGreen}{rgb}{0.0, 0.27, 0.13}
\definecolor{Fuchsia}{rgb}{1.0, 0.0, 1.0}

\definecolor{amethyst}{rgb}{0.6, 0.4, 0.8}
\definecolor{aureolin}{rgb}{0.99, 0.93, 0.0}
\definecolor{parisgreen}{rgb}{0.31, 0.78, 0.47}
\definecolor{persianred}{rgb}{0.8, 0.2, 0.2}

\def\mclearpage{}

\newcommand{\appref}[1]{the appendix \OldS~#1}
\newcommand{\apprefp}[1]{appendix \OldS~#1}
\newcommand{\apprefs}[2]{the appendix \OldS~#1 and #2}

\def\secUseCaseSpecApp{1}

\def\secSemanticsApp{2.1}

\def\secComSubexpElim{2.2.1}
\def\secExtAppUnaryLiteral{2.2.3}
\def\secVertexVarApp{2.2.4}

\def\secMultipleRounds{2.2.6}

\def\secExampleFusions{2.3}

\def\secIterRedApp{3.1}

\def\secPushIterCorrectness{3.1.2}
\def\secAsynchronous{3.1.3}
\def\secStreamingGraphs{3.1.4}
\def\secFactoredPathBasedRed{3.1.5}

\def\secSynthesisApp{3.2}

\def\secCompProof{4.2}
\def\secFutSoundness{4.3}

\def\secProofs{4.4}

\def\secFrameworksAll{5.1}

\def\secEvaluationFusionScale{6.1}
\def\secEvaluationFusionExec{6.2}

\begin{document}



\title[]{\name: Graph Analytics Fusion and Synthesis}



 \author{Farzin Houshmand}

 \affiliation{
   \institution{University of California, Riverside}           
   \country{USA}                   
 }
 \email{fhous001@cs.ucr.edu}         


 \author{Mohsen Lesani}
 \affiliation{
   \institution{University of California, Riverside}           
   \country{USA}                   
 }
 \email{lesani@cs.ucr.edu}         

\author{Keval Vora}
\affiliation{
\institution{Simon Fraser University}           
\country{Canada}                   
}
\email{keval@sfu.ca}         

\begin{abstract}

Graph analytics elicits insights from large graphs to inform 
critical decisions for
business, safety and security.
Several large-scale graph processing frameworks
feature efficient runtime systems;
however, they often provide programming models 
that are low-level and subtly different 
from each other.
Therefore, 
end users can find
implementation and specially optimization of
graph analytics time-consuming and error-prone.
This paper 
regards the abstract interface of the graph processing frameworks
as the instruction set for graph analytics,
and
presents \name,
a high-level declarative specification language for graph analytics
and 
a synthesizer that automatically generates efficient code for 
five
high-performance graph processing frameworks.
It features novel semantics-preserving fusion transformations
that optimize
the specifications and
reduce them 
to three primitives:
reduction over paths, 
mapping over vertices
and
reduction over vertices.
%
Reductions over paths are commonly calculated based on push or pull models
that iteratively apply kernel functions at the vertices.
This paper presents 
conditions, parametric in terms of the kernel functions,
for the correctness and termination
of the iterative models,
and 
uses these conditions as specifications 
to automatically synthesize
the kernel functions.
%
%
Experimental results show that 
the generated code 
matches or outperforms
hand-optimized code,
and
that fusion 
accelerates execution.

\end{abstract}

\keywords{}


\maketitle


\section{Introduction}

Large-scale \emph{graph analytics} has recently gained popularity due to its growing applicability across various important domains including social networks, market influencer analysis, bioinformatics, criminology, and machine learning and data mining. 
Several large-scale graph processing
systems~\cite{powergraph,ligra,gemini,gridgraph,pregel,xstream,graphit} have been developed to enable efficient graph analysis across shared memory and distributed platforms. 
%
The
provided programming models
require graph analysis problems to be expressed in terms of 
local 
kernel functions over
vertices and edges.
However, graph analyses are best 
expressed using \emph{higher-level abstractions} such as reduction over 
paths.
For instance, shortest path, reachability and connected component problems are fundamentally formulated in terms of paths. 
%
%
Further, 
elaborate 
graph analysis problems 
that involve multiple 
reductions over paths or vertices
are 
difficult
to correctly implement
using the offered low-level programming models.
%
%
%
More importantly, 
manual \emph{optimizations}
such as 
merging multiple iterations
can be time-consuming and error-prone.
In particular, reasoning about the 
\emph{correctness and termination} properties 
requires
challenging analysis on 
the values of vertices
across iterations 
that emulate values for paths.

%

This project regards the interface of the graph processing frameworks
as the instruction set for graph analytics,
and introduces \name, 
a graph analytics language and synthesizer.
The \name\ language
is a \emph{high-level declarative specification language}
that provides features for common graph processing idioms
such as reduction over paths,
and mapping and reduction over vertices.
We show that it can 
easily and
concisely capture
the common 
graph analysis problems.
Given a specification,
the \name\ synthesizer
\emph{automatically synthesizes code} for 
five
graph processing frameworks:
Ligra~\cite{ligra},
GridGraph~\cite{gridgraph},
PowerGraph~\cite{powergraph},
Gemini~\cite{gemini},
and
GraphIt \cite{graphit}.

To synthesize efficient implementations,
\name\ 
optimizes the specification by syntactic \emph{fusion transformations}
that fuse similar operations to be executed together.
%
We formalize the syntax and the semantics of the \name\ language
and the fusion rules,
and prove that fusion is semantics-preserving.
As described above,
fusion reduces specifications to
the sequence of three primitives:
reduction over paths,
mapping over vertices
and
reduction over vertices.

Graph analytics frameworks offer
\emph{iterative programming models} 
to calculate reduction over paths.
%
%
The values for vertices or edges are calculated
iteratively based on the values of neighbors.
Influenced by their runtime systems,
these frameworks differ on how values are propagated between iterations.
For example, PowerGraph~\cite{powergraph} allows computations to 
pull
in values from incoming neighbors or to 
push
out values to outgoing neighbors, 
whereas Ligra~\cite{ligra} and GridGraph~\cite{gridgraph} only allow pushing values.
Further, Gemini~\cite{gemini} requires 
both pull-based
and push-based 
implementations of the computation
so that it can dynamically switch between the two to maximize performance. 
%
%
Not only the propagation methods, 
but also \emph{system-specific} nuances 
make 
the implementation of the same analysis problem
subtly different from a framework to another.
For example, a push-based implementation in Ligra requires an atomic function that operates over a single edge whereas Gemini requires two functions: the first determines the value to be pushed from the source, and the second operates directly over the outgoing edges, and updates the target values atomically. 
%

We formalize 
several \emph{iteration models}
that given certain kernel functions,
calculate path-based reductions.
%
For each iteration model,
we present \emph{correctness and termination conditions}
for
candidate kernel functions.
%
Given 
a path-based reduction,
\name\ synthesizer 
enumerates candidate kernel functions
and
uses the correctness conditions as specifications
to
automatically \emph{synthesize
the kernel functions}.
Fusion reduces specifications to
reduction over paths,
mapping over vertices
and
reduction over vertices.
Subsequently, the synthesizer reduces path-based reductions to iterative calculations.
Thus,
graph analysis is reduced to
\emph{iteration-map-reduce primitives}.
We show how each of these primitives can be immediately implemented in each of the 
five
target frameworks.

We apply \name\ to 
common graph analysis use-cases and 
generate code for each of the
five
frameworks.
%
The experimental results show 
that synthesized programs are equally or more efficient than 
hand-optimized programs,
and
that fusion significantly reduces execution time.

In summary, this paper makes the following contributions:
%
(1)
The graph analytics specification language \name\ and its semantics 
(\autoref{sec:overview} and \autoref{sec:lang-spec}),
%
(2)
Semantics-preserving 
and platform-independent
fusion transformations to optimize graph analytics
(\autoref{sec:fusion}),
%
(3)
The formalization of iterative graph computation models
(\autoref{sec:iterative-models}),
their correctness and termination conditions
(\autoref{sec:iter-reduction}),
and synthesis of their kernel functions
(\autoref{sec:synthesis}),
and
%
(4)
The \name\ 
synthesis tool
that generates code for
five
graph processing frameworks
and its experimental results
(\autoref{sec:experiments}).


\co{
    Background, importance
    Language 
        Higher level abstraction.
        Uniformity, single high-level language.
    Fusion, optimization
    Correctness and synthesis
    Implementation, Iteration-Map-Reduce, Mapping to four frameworks
    Evaluation synthesized versus handwritten, fusion
}

\mclearpage
\section{Overview}
\label{sec:overview}

We start with an overview.
We first
present the \name \ specification language through examples,
and then show how specifications can be fused to equivalent more efficient specifications.
Then, we illustrate 
iterative reductions
and present a glimpse of their correctness conditions
and how the kernel functions can be synthesized.

 \begin{wrapfigure}{R}{0.6\textwidth}  
\small
\resizebox{\linewidth}{!}{
$
      \begin{array}{cc}
        
        \begin{array}{llll}
        \rulelabel{SSSP}
        \SP(s)(v)& = & 
           \displaystyle \min_{  p \in \Paths(s, v)} \weight(p)
           &
            \\[8pt]


            \rulelabel{CC}
            \CC(v) & = & 
            \displaystyle \min_{p \in \Paths(v)} \head(p)
            &
            \\[8pt]

            \rulelabel{BFS}
            \BFS(s)(v)& = & 
           \penultimate(\displaystyle \argmin_{ p \in \Paths(s,v)} \length(p))
           &
            \\[8pt]

            \rulelabel{WSP}
            \WSP(s)(v) &=&
            \llet P \coloneqq
            \displaystyle \argsmin_{p \in \Paths(s, v)} \length(p)
            \lin            
            \\ &&
            \displaystyle 
                \max_{p \in P}
                    \ 
                    \capacity(p)
            \\[8pt]

            \rulelabel{NSP}
            \NSP(s)(v)& = & 
            \left| \, \displaystyle \argsmin_{p \in \Paths(s, v)} \weight(p) \, \right|
            &
            \\[12pt]

            \rulelabel{NWR}            
            \NWR(s)(v)& = & 
            \frac{
                \displaystyle 
                \min_{p \in \Paths(s, v)} \capacity(p) 
            }{
                \displaystyle 
                \max_{p \in \Paths(s, v)} \capacity(p) 
            }
            &
            \\[8pt]

            \\
            \rulelabel{Trust}
            \Trust(v) & = &
            \displaystyle
            \max_{s \in \overline{s}}
            \displaystyle
            \left(
            \frac{
                \displaystyle 
                \max_{p \in \Paths(s, v)} \capacity(p) 
            }{
                \displaystyle 
                \min_{p \in \Paths(s, v)} \length(p) 
            }
            \right)
            &
            \\[8pt]


           \\
            \rulelabel{Radius}            
            \Radius & = & 
            \displaystyle \min_{s \in \{\overline{s}\}}
            \displaystyle \max_{v \in \Vertices} 
            \min_{ p \in \Paths(s,v)} \length(p)
            &
            \\[8pt]

            \\
            \rulelabel{DRR}
            \DRR & = &
            \displaystyle
            \frac{
                \displaystyle \max_{s \in \{\overline{v}\}} 
                \displaystyle \max_{v \in \Vertices} 
                \min_{ p \in \Paths(s,v)} \length(p)
            }{
                \displaystyle \min_{s \in \{\overline{v}\}} 
                \displaystyle \max_{v \in \Vertices} 
                \min_{ p \in \Paths(s,v)} \length(p)
            }
            &
            \\[12pt]


            \rulelabel{DS}
            \DS(s) & = &
            \displaystyle
            \bigcup_{
                v \in \Vertices \, \land \,
                \displaystyle
                \left(
                \min_{  p \in \Paths(s, v) }
                \weight(p)
                \right)
                > 7
            }
            \{v\}
            &
            \\[8pt]

            \\
            \rulelabel{RDS}
            \RDS(s) 
            & = &
            \llet 
                \SP \coloneqq \lambda s, v. \ 
                \displaystyle \min_{  p \in \Paths(s, v) } \weight(p) 
            \lin
            \\
            & &
            \llet 
                \WP \coloneqq \lambda s, v. \ 
                \displaystyle \max_{  p \in \Paths(s, v) } \capacity(p)
            \lin
            \\
            & &
            \displaystyle
            \min_{
                v \in \Vertices \, \land \,
                \displaystyle
                \SP(s, v)
                < 
                \Radius
            }
            \WP(s, v)
            &

         \end{array}
      \end{array}
   $
   }
   
   \caption{\label{fig:benchs-spec}A subset of 
        use-cases in \name. 
    }
\end{wrapfigure}

\textbf{Graph Analysis Specification. \ }
%
The \name \  language
declaratively and
concisely
captures
\emph{mathematical specifications of 
graph analysis computations}.
The language design is guided by common idioms in graph processing use-cases.
\name \  supports 
reduction over values of paths to a vertex 
and also 
mapping and reduction over values of vertices.
We present example specifications
in \autoref{fig:benchs-spec}.
More use-cases are available in 
\appref{\secUseCaseSpecApp} \cite{appendix}.


The use-case \irule{SSSP} specifies the weight of
the shortest path from the source vertex $s$ to each vertex $v$.
The set of paths from a source vertex $s$ to a destination vertex $v$ is denoted by $\Paths(s, v)$.
The specification applies the minimum reduction function $\min$ to 
the result of applying the weight function $\weight$
to all paths $p$ in 
$\Paths(s, v)$.
%
%
The specification of connected component (for undirected graphs) \irule{CC}
takes the smallest identifier of the vertices in a component as
the representative identifier of that component.
The set of all paths (from any source vertex) to a destination vertex $v$ is denoted by $\Paths(v)$.
The specification \irule{CC} defines
the connected component of each vertex $v$
as the minimum identifier of the head vertices of the 
paths $\Paths(v)$.
%
The above two specifications apply a reduction function $\R$ 
to the result of a path function $\F$
for 
a set of paths.
We call these reductions \emph{path-based reductions}.
Similarly,
the breadth-first-search use-case \irule{BFS}
finds
the parent of each vertex in the breadth-first-search tree.
For each vertex $v$,
it 
specifies a path-based reduction to find
the shortest-length path from the source $s$ to the vertex $v$
and
returns the penultimate of that path.
The penultimate of a path is the vertex before the last in the path.
%
%
%
The specification uses the reduction function $\argmin$ to get the path with the minimum length rather than the minimum length itself,
and then applies the $\penultimate$ function to the path.
A simpler specification can simply apply $\min$ and return the minimum path length.

Path-based reductions 
can be nested.
The use-case \irule{WSP} specifies the widest shortest path from a source $s$ to each vertex $v$.
We use the $\llet$ syntactic sugar to enhance readability.
\irule{WSP}
has a nested reduction
(with the reduction function $\argsmin$)
to find the shortest paths, and
then a nesting reduction
to find the widest capacity in those paths. 
\irule{WSP} is used as a metric of the trust of a user to other users in social networks
where 
the 
capacity of each edge is the local trust rating of the source user to the sink user
\cite{golbeck2005computing}.
Intuitively,
users with
wider (stronger trust ratings)
and 
shorter (closer)
paths 
are 
more trustworthy
sources of information.
Similarly, 
the use-case \irule{NSP} specifies the number of shortest paths from a source $s$ to each vertex $v$.
It uses a nested reduction 
to find the shortest paths and
then applies the cardinality operator to the resulting set.
(We will see in \autoref{sec:extensions} that
cardinality is a syntactic sugar for a path-based reduction with the sum $\sum$ 
function.)
Mathematical operators can be applied to path-based reductions.
The use-case \irule{NWR} specifies 
the narrowest to widest path ratio
from a source to each vertex.
It divides two path-based reductions.
Similarly, 
the use-case \irule{Trust}
is the result of division and maximum operations between path-based reductions.
It specifies the trust from a set of users $\{\overline{s}\}$ to each other user.
As before, wider and shorter paths are favored.

The values of vertices 
calculated by a path-based reduction
can be subsequently reduced by a \emph{vertex-based reduction}.
%
%
The \irule{Radius} use-case specifies the radius of the graph by
sampling the eccentricity of a set of sources $\{\overline{s}\}$.
A vertex-based reduction with the reduction function $\max$
finds the longest of the shortest paths 
over all vertices.
Similar to path-based reductions,
mathematical operators can be applied to vertex-based reductions.
As the set of sampled sources is finite,
the outer $\min$ function can be unrolled to an infix operator between vertex-based reductions.
Similarly,
the use-case \irule{DRR},
that is the ratio of the the diameter over the radius of the graph,
is specified as division, maximum and minimum operations between vertex-based reductions.


The use-case \irule{DS}
specifies the set of vertices with the distance of at least 7 from the source $s$.
%
The union $\union$ vertex-based reduction is used to calculate the set.
The set of vertices that it is applied to are constrained
by a nested path-based reduction to specify the distance.
%
%
(In 
\autoref{sec:extensions},
we show that
constrained vertex-based reductions can be
desugared to standard vertex-based reductions 
that are applied to
path-based reductions on pairs of values.)
Similarly, the use-case \irule{RDS}
is specified as a constrained vertex-based reduction.
Given a source $s$,
it calculates
the narrowest of the
widest paths
to vertices
within the 
radius of $s$ ($k$-hop neighbourhood of $s$ where $k$ is the radius of the graph).
%
In a social network,
\irule{RDS} can represent a measure of the least amount of 
trust from a user to her neighbourhood.

\begin{figure}
\small


\begin{align}
%
%
%
\Radius \ 
& =
\displaystyle \min_{s \in \{s_1, s_2\}} 
\displaystyle \max_{v \in \Vertices} 
\min_{ p \in \Paths(s,v)} \length(p)
\label{eq:r1}
%
%
%
\\ & = 
\min \left(
\displaystyle \max_{v \in \Vertices} 
\min_{ p \in \Paths(s_1, v)} \length(p), \ \
\displaystyle \max_{v \in \Vertices}
\min_{ p \in \Paths(s_2, v)} \length(p)
\right)
\label{eq:r2}
\\ & = 
%
%
%
%
%
%
%
\min 
\left(
\left(
{\begin{array}{l}
\ilet x \coloneqq \displaystyle \min_{s_1} \ \length \lin
\\ 
\mlet x' \coloneqq x \lin
\\ 
\rlet x'' \coloneqq 
\max \ x' 
\lin
\\
x''
\end{array}}
\right),
\ \ 
\left(
{\begin{array}{l}
\ilet y \coloneqq \displaystyle \min_{s_2} \ \length \lin
\\ 
\mlet y' \coloneqq y \lin
\\ 
\rlet y'' \coloneqq 
\max \ y' 
\lin
\\
y''
\end{array}}
\right)
\right)
\label{eq:r4}
\\ & =
%
%
%
%
\left(
{\begin{array}{l}
\ilet \<x, y\> \coloneqq 
    \< \displaystyle \min_{s_1} \ \length, \ 
        \displaystyle \min_{s_2} \ \length \> \lin
\\ 
\mlet \<x', y'\> \coloneqq \<x, y\> \lin
\\ 
\rlet \<x'', y''\> \coloneqq 
    \< \max \ x', \ 
        \max \ y' \>
    \lin
\\
\min(x'', y'')
\end{array}}
\right)
\label{eq:r5}
\\ & =
%
%
%
%
\left(
\begin{array}{l}
\ilet \<x, y\> \coloneqq 
    \underset{\<s_1, s_2\>}{\R} \ \F
    \lin
\\ 
\mlet \left\< x', y' \right\> \coloneqq \<x, y\> \lin
\\ 
\rlet \<x'', y''\> \coloneqq 
    \left\< \max \ x', \ 
        \max \ y' \right\>
    \lin
\\
\min(x'', y'')
\end{array}
\right)
\mbox{\ \ where \ \ }
\begin{array}{l}
\F \coloneqq \lambda p. \ \left\< \length(p), \length(p) \right\> 
\\
\R \, ( \<a, b\>, \<a', b'\> ) \coloneqq
\\ \t
\left\<
\min (a, a'), \min (b,  b') 
\right\>
\end{array}
\label{eq:r6}
\\ & =
%
%
%
%
\left(
\begin{array}{l}
\ilet \<x, y\> \coloneqq 
    \underset{\<s_1, s_2\>}{\R} \ \F
    \lin
\\ 
\mlet \<x', y'\> \coloneqq \<x, y\> \lin
\\ 
\rlet \<x'', y''\> \coloneqq 
    \R' \ \< x',  y' \>
    \lin
\\
\min(x'', y'')
\end{array}
\right)
\mbox{\ \ where \ \ }
\begin{array}{l}
\R' \, ( \<a, b\>, \<a', b'\> ) \coloneqq
\\ \t
\<\max (a, a'), \max (b,  b') \>
\end{array}
\label{eq:r7}
\end{align}

\caption{\label{fig:radius-fusion}Fusion of the \irule{Radius} use-case.}
\end{figure}

\textbf{Fusion. \ }
A naive execution of specifications may execute 
path-based and vertex-based reductions
multiple times.
We show that 
multiple such reductions can be fused into a single reduction
and 
represented as a common triple-let form
with separate terms for 
path-based reduction,
mapping over vertices 
and 
vertex-based reduction.
%
The fused computation can execute significantly faster.
%

For example, 
the \irule{Radius} use-case that we saw in \autoref{fig:benchs-spec}
includes multiple path-based reductions one per source that can be fused together.
Further,
the path-based reductions are enclosed by vertex-based reductions
that can be fused together as well.
%
We illustrate this fusion
in \autoref{fig:radius-fusion}.
For simplicity, we consider sampling for two sources $\{s_1, s_2\}$.
We consider the fusion steps in turn.
The specification of \irule{Radius} is represented in \autoref{eq:r1}.
In \autoref{eq:r2},
the outer $\min$ function over the two sources is unrolled.
In \autoref{eq:r4},
we restate 
each of the two reductions
in a \emph{triple-let} form.
\name \  features a triple-let term that
separates path-based reductions, 
mapping over vertices
and vertex-based reductions,
and
thus facilitates fusion.
%
%
The term
$\displaystyle \max_{v \in \Vertices} 
\min_{ p \in \Paths(s_1, v)} \length(p)$
is rewritten as
the following three lets.
The first let,
$\ilet x \coloneqq \displaystyle \min_{s_1} \ \length$,
calculates a path-based reduction.
For each vertex,
it calculates the weight of
the shortest path from the source $s_1$
and
binds the result to $x$.
The second let applies a map function
on the results of the path-based reductions
in each vertex.
In this case, there is only one path-based reduction;
therefore, the map function in the second let,
$\mlet x' \coloneqq x$,
is simply the identity function.
%
%
(In use-cases that the nesting vertex-based reduction is applied to
an expression over path-based reductions,
the map in the second let captures the expression.)
%
%
After fusion,
the second let applies division as the map function.)
%
The third let 
calculates a reduction over all vertices.
In this example,
$\rlet x'' \coloneqq \max \ x'$
calculates the maximum value of over all vertices
and binds the result to $x''$.

Next,
in \autoref{eq:r5},
the two triple-let terms are fused into one
by pairing the operations of the corresponding lets.
The 
outer $\min$ is applied to the two final results $x''$ and $y''$.
In the next two steps, 
the paired path-based and vertex-based reductions are fused.
In \autoref{eq:r6},
the two path-based reductions of the first let are fused into one.
The two sources $s_1$ and $s_2$
are 
used
to initialize the first and second elements of the pairs respectively.
The fused reduction calculates the pair of the two values simultaneously.
The fused path function $\F$ returns the pair of 
the results of the two path functions.
Similarly, the fused reduction function $\R$
applies the two reduction functions
to the first and second elements of the input pairs respectively.
Finally, in \autoref{eq:r7},
the pair of vertex-based reductions of the third let are fused into one.
%
The fused reduction function $\R'$
applies the two reduction functions
to the first and second elements of the input pairs respectively.
This simple example showcased fusion.
We formally present the complete set of fusion rules in \autoref{sec:fusion}.

The final term 
represents
the original specification of \irule{Radius} as
an equivalent sequence of 
one path-based reduction,
one map in each vertex,
and
one reduction over all vertices.
Path-based reduction are calculated iteratively.
Thus, 
fusion reduces \name \  specifications to 
three primitives:
\emph{Iteration-Map-Reduce}:
iteration for 
iterative path-based reduction, 
map for mapping over vertices 
and
reduce for reduction over vertices.
Map and reduce over vertices can be directly implemented;
we now discuss iterative path-based reduction.


 \begin{wrapfigure}[17]{R}{0.5\textwidth}

\footnotesize
\centering
\begin{subfigure}{0.48\textwidth}
\centering
\includegraphics[scale=0.35]{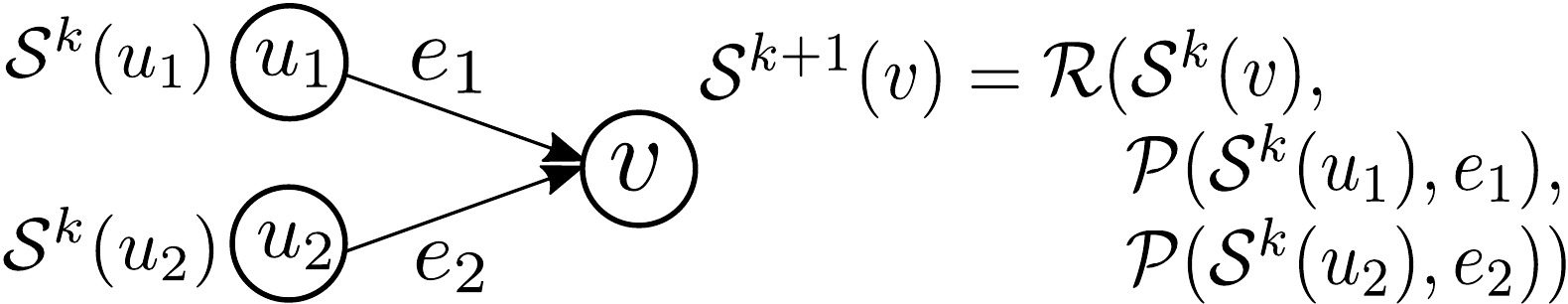}
\caption{Propagation and update in the pull model}
\label{fig:overview-a}
\end{subfigure}

 \vspace{.1cm}
\begin{subfigure}{0.48\textwidth}
\centering
\includegraphics[scale=0.35]{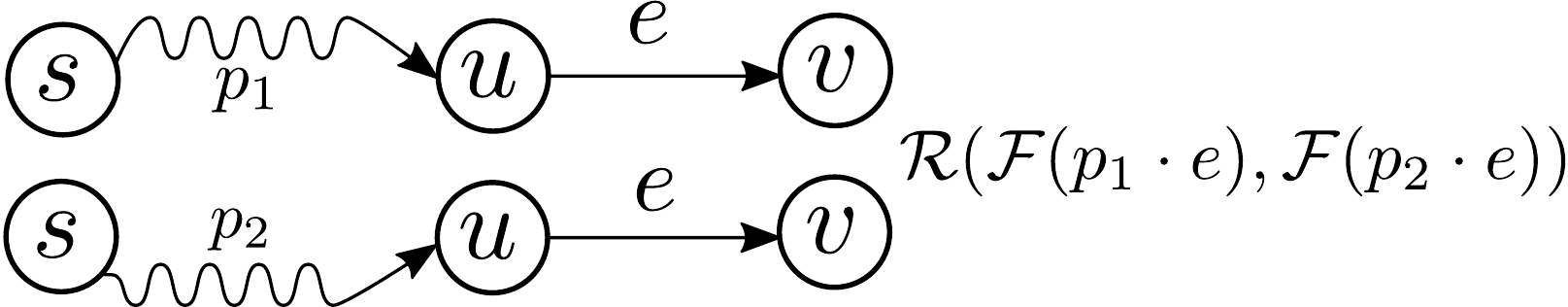}
\caption{Separate calculation for paths and then reduction}
\label{fig:overview-b}
\end{subfigure}
%

%
\begin{subfigure}{0.48\textwidth}
\centering
\includegraphics[scale=0.35]{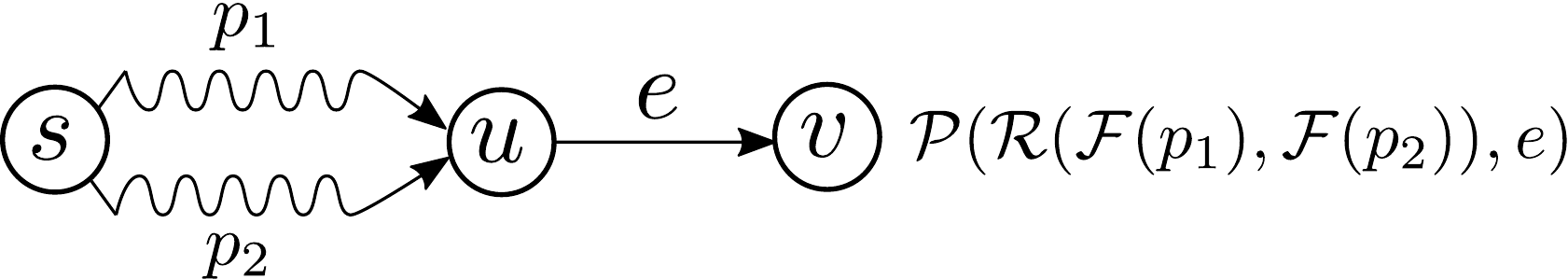}
\caption{Reduction at predecessor and then propagation}
\label{fig:overview-c}
\end{subfigure}

\caption{The pull model and its correctness. The path $p \cdot e$ denotes the extension of 
path $p$ with 
edge $e$.}
\label{fig:overview}
\end{wrapfigure}

\textbf{Iteration. \ }
Calculating path-based reductions
by
explicit enumeration of paths is prohibitively inefficient.
Instead, path-based reductions are calculated iteratively 
by local updates on the value of vertices
based on the values of their neighbors.
%
As an example, we consider the pull model for idempotent reduction functions.
Let us consider the simple shortest path use-case \irule{SSSP}
that we saw in \autoref{fig:benchs-spec}.
It specifies 
a path-based reduction from the source $s$
where
the reduction function $\R$ is $\min$
and 
the path function $\F$ is $\length$.

Each vertex stores a value; 
we denote the value of a vertex $v$ in the iteration $k$ as $\S^k(v)$.
(The fused path-based reduction of the \irule{Radius} use-case 
that has two sources stores a pair of values.)
The iterative calculation is based on 
the initialization function $\I$,
the propagation function $\P$
and
the reduction function $\R$.
The initialization function $\I$ is a function from vertices to their initial value.
For the \irule{SSSP} use-case, 
$\I$ is $\lambda v. \ \lif \ (v = s) \ 0 \ \lelse \ \bot$
that initializes the value of the source $s$ to zero (the some value of zero to be more precise)
and
the other vertices to none $\bot$.
In each iteration,
if the value of a vertex changes, 
its successors are notified to be active in the next iteration.
As \autoref{fig:overview-a} shows,
in an iteration $k + 1$,
an active vertex $v$ pulls the value $\S^k(u)$ of each of its predecessors $u$.
It then applies the propagation function $\P$ to 
the value $\S^k(u)$ and the edge $\<u, v\>$.
%
%
It then reduces 
using $\R$
the results of propagation form the predecessors
together and
with 
the current value $\S^k(v)$ of $v$
to calculate the new value $\S^{k+1}(v)$ of $v$.
After a number of iterations,
the values of 
the vertices converge.
The calculation stops when
the values of all vertices stay unchanged
in two consecutive iterations.

\textbf{Correctness and Synthesis. \ }
%
In \autoref{sec:iter-reduction},
for a given specification,
we formalize \emph{correctness and termination conditions}
that are parametric in terms of the
candidate initialization and propagation functions.
We present sufficient conditions for a comprehensive set of iteration methods.
%
As an example, we consider the pull model
and
illustrate one of the correctness conditions
on the propagation function $\P$
in 
\autoref{fig:overview-b} and 
\autoref{fig:overview-c}.
Consider a vertex $v$ and a predecessor $u$ of $v$.
Consider the reduction over all the paths to $v$ that go through $u$.
\autoref{fig:overview-b} shows
the direct calculation where
the value of the path function for each path to $v$ is separately calculated
and
then the results are reduced.
On the other hand,
\autoref{fig:overview-c} shows
a calculation using the propagation function $\P$ where
first, 
the values of the path function for the paths to the predecessor $u$ are calculated
and
reduced,
and then,
the result is propagated to $v$.
%
%
In order to correctly calculate the path-based specifications,
the iterative computation requires
the result of the above two calculations to be the same.
Global reductions over paths
should be equivalent to
local propagations from predecessors.
%
%
Further, to reason about termination,
we formalize the termination conditions for the two iteration models
in \autoref{sec:iter-reduction}.
Iterations incrementally consider longer paths.
Cycles of a graph generate an infinite number of paths
and can cause divergence.
%
However, sometimes 
adding longer paths has no effect on the result of the reduction.
For example for the shortest path use-case \irule{SSSP} (with non-negative edges),
after a certain number of iterations,
all the simple paths of the graph are already considered,
and longer cyclic paths cannot 
    improve the shortest path.


%
%
%
In \autoref{sec:synthesis},
we use the correctness conditions to
synthesize correct iteration functions $\I$ and $\P$.
In particular,
we 
apply 
type-guided
enumerative synthesis
to find candidates
and
automatic solvers to
check the validity of the correctness conditions for each.
The result is 
correct-by-construction
kernel functions
that can iteratively calculate 
path-based reductions.
%
In \autoref{sec:frameworks},
we use the synthesized iteration functions to
generate 
code for 
five
high-performance graph computation frameworks.
%

\section{Iterative Models}
\label{sec:iterative-models}

We formalize four canonical models for iterative graph computations:
the pull and push models with idempotent and non-idempotent reduction.
Graph computation frameworks~\cite{powergraph,ligra,gemini,gridgraph,pregel,xstream}
implement variants of these models.
Later in \autoref{sec:mapping},
we use them
to implement path-based reductions and 
the correctness conditions of these implementations.

In these models, 
each vertex is first initialized.
Then,
the value of each vertex
is iteratively updated
based on the values of its predecessors.
In each iteration,
the values of the predecessors are pulled or each predecessor pushes its value to the vertex.
Then, 
the values of the predecessors are reduced to a single value.
Before assigning
the reduced value
to the vertex,
a final function may be applied to it.
The iteration stops when the value of no vertex changes.

The iteration models are parametrized by four kernel functions:
$\I$, $\P$, 
$\R$ and $\E$.
The initialization function $\I$ defines the initial value for each vertex.
The propagation function $\P$, given a value $n$ and an edge $e$
where $n$ is the current value of the source of $e$,
defines the value that is propagated to the destination of $e$.
%
The reduction function $\R$ defines how the propagated values are aggregated.
The epilogue function $\E$,
given an aggregated value $n$,
defines the final update to $n$.

This work presents a
high-level language
to specify the kernel functions.
It compiles kernels specified in this language to 
executable programs in 
five
graph processing frameworks.
The language grammar for bodies of the kernel functions
is presented in 
\autoref{fig:iter-kern-exam}a.
Later in \autoref{sec:synthesis},
the same grammar is used 
by
the synthesis process;
given higher-level specifications,
it 
automatically generates the kernel functions in this language.
%
\autoref{fig:iter-kern-exam}b shows
the iterative kernel functions for
two use-cases:
the shortest path \irule{SSSP}
and
the page-rank \textsc{PageRank}.
%
For the shortest path \irule{SSSP} use-case,
the initialization function $\I$
initializes the source vertex $s$ to $0$ and the other vertices to none $\bot$.
The propagation function $\P$ 
adds the value $n$ of the predecessor with the weight of the edge $e$.
The reduction function is $\R$ is the minimum (that is idempotent)
and the epilogue function is the identity function.
For the page-rank \textsc{PageRank} use-case,
the initialization function $\I$
divides the value $1$ between the number of vertices $|V|$.
The propagation function $\P$ 
divides the value $n$ of the predecessors between its successors.
The reduction function is $\R$ is sum (non-idempotent).
The epilogue function 
multiplies the sum with the damping factor $\gamma$ and adds a constant.
We now consider each model.

\begin{figure*}
\small

$
\begin{array}{cc}

\begin{array}{rcl@{\qquad}l}
      e
         & ::= &
            n
            \ \ | \ \
            v
        &
            \!\!\!\!\!\!
            \mbox{Body Exp}

    \\    
         & | &
            e+e
            \ \ | \ \
            e-e
            \ \ | \ \
            -e
            \ \ | \ \
            \<e, e\>
    \\
         & | &
            e*e
            \ \ | \ \
            e \, / \, e
            
            \ \ | \ \
            e = e
            \ \ | \ \
            e < e
         &

    \\
         & | &
            \min(e,e)
            \ \ | \ \
            \max(e,e)

    \\
         & | &
            \lif \ (e) \lthen e \ \lelse \ e
         &
            \!\!\!\!\!\!
    \\
         & | &
            \mathsf{weight}(e)
            \ \ | \ \
            \mathsf{capacity}(e)
            
    \\
    &| &
            \mathsf{indeg}(e)
            \ \ | \ \
            \mathsf{outdeg}(e)
            
    \\
         & | &
            \mathsf{src}(e)
            \ \ | \ \
            \mathsf{dst}(e)
         &
         \!\!\!\!\!\!
    \\
         & | &
        |V|
        &
        \!\!\!\!\!\!
        \mbox{Graph Order}
    \\    
         n
         & ::= &
            0
            \ \ | \ \
            1
            \ \ | \ \
            ..
            \ \ | \ \
            \True
            \ \ | \ \
            \False
         &
         \!\!\!\!\!\!
         \mbox{Literal}
    \\    
         v
         &  &
         &
         \!\!\!\!\!\!
         \mbox{Variable}
      \\   
      \multicolumn{4}{c}{\mbox{(a) Grammar} }
   \end{array}

&

\begin{array}{l}
\begin{array}{rclr}
    \multicolumn{3}{l}{\textsc{SSSP}}
    \\
    \ \ \ \
    \I & \coloneqq &
        \lambda v. \ \ 
        \lif \ (v = s) \ 0 \ \lelse \ \bot
    \\
    \P & \coloneqq &
        \lambda n, e. \ \         
        n + \weight(e)
    \\
    \R & \coloneqq &
        \lambda v, v'. \ \
        \min(v, v')
    \\
    \E & \coloneqq &
        \lambda n. \ \ 
        n
\end{array}

\\

\begin{array}{rclr}
    \multicolumn{3}{l}{\rulelabel{PR} \textsc{PageRank}} 
    \\
    \ \ \ \ 
    \I & \coloneqq &
        \lambda v. \ \ 
        1 \, / \, |V|
    \\
    \P & \coloneqq &
        \lambda n, e. \ \ 
        n \, / \, \mathsf{outdeg}(\mathsf{src}(e))
    \\
    \R & \coloneqq &
        \lambda v, v'. \ \ 
        v + v'
    \\
    \E & \coloneqq &
        \lambda n. \ \ 
        \gamma * n \,\, + \,\, (1-\gamma) \, / \, |V|
\end{array}
\\
\\
\multicolumn{1}{c}{\mbox{(b) Examples}}
\end{array}

\end{array}
$

\caption{
    (a) Grammar for kernel functions
    (b) Example kernel functions.
    ($\min$ and $+$ filter none values $\bot$.)
}
\label{fig:iter-kern-exam}
\end{figure*}

\textbf{Pull Model. \ }
The characteristic of the pull model is that
vertices pull the values of their predecessors
to calculate their new values.
We consider the pull model for idempotent and non-idempotent reduction functions in turn.

\textit{Pull model with idempotent reduction ($\pulli$). \ }
The pull model for idempotent reduction is represented in 
\autoref{fig:pull-push-iter},
\autoref{def:computation-pull}.
The value of the vertex $v$ in the iteration $k$ is represented by $\S_\pulli^k(v)$.
In the beginning $k = 0$,
the vertices have no value $\bot$.
In the first iteration $k = 1$, 
they are initialized by the initialization function $\I$.
In subsequent iterations $k+1, \ k \geq 1$,
each vertex $v$ pulls values of its predecessors. 
For each predecessor $u$,
the propagation function $\P$ is applied
to the value of $u$ (from the previous iteration $k$)
and the connecting edge $\<u, v\>$.
Then, 
as illustrated in \autoref{fig:overview-a},
all the propagated values are reduced by $\R$
with each other
and 
then 
with the previous value of $v$.
Finally, the application of the epilogue function to the reduced value 
results in the new value of $v$.
%
As an optimization, 
the above update is performed only 
if the value of at least one predecessor has been updated in the previous iteration.

\textit{Pull model with non-idempotent reduction ($\pulln$). \ }
The pull model for non-idempotent reduction 
is represented in
\autoref{fig:pull-push-iter},
\autoref{def:computation-pull-noni}.
%
%
The value of the vertex $v$ in the iteration $k$ is represented as $\S_\pulln^k(v)$.
Similar to the pull model for idempotent reduction,
the 
values from predecessors are propagated and reduced.
The value that 
each vertex stores is a reduction of a set of values.
Consider a vertex $v$ a predecessor $u$ of $v$.
Assume that  the value of $u$ represents the reduction of a set $S$ of elements.
After the value of $u$ is propagated to $v$,
the value of $v$ includes the reduction of $S$.
Assume that 
the value of $u$ is updated to represent the reduction more elements.
%
Since the reduction is non-idempotent,
propagating the new value of $u$ to $v$ and reducing it with the current value of $v$
results in the two times reduction of $S$ in $v$.
Therefore, 
to avoid this duplicated reduction,
the difference of this model with the previous model 
is that 
after 
reducing the propagated values,
the result
is not reduced with the previous value of the vertex.

%

\begin{figure*}
\small

$
\begin{array}{l}
\CPreds^k(v) = 
\left\{ u \ | \ u \in \preds(v) \land \S^k(u) \neq \S^{k-1}(u) \right\}
\end{array}
$

\begin{definition}[Pull (idempotent reduction)]
\label{def:computation-pull}
$$
\begin{array}{ll}
\S_\pulli^0(v) \coloneqq \bot
\ \\
\S_\pulli^1(v) \coloneqq \I(v)
\ \\
\S_\pulli^{k+1}(v) \coloneqq
\twopartdef{
    \S_\pulli^k(v)
}{
    \CPreds^k(v) = \emptyset
}{
    \E \left[ 
    \R \left(
        \S_\pulli^k(v), \ 
        \undersetp{ \, u \in \preds(v) }{\R}
        \, \P \! \left( \S_\pulli^k(u), \<u, v\> \right)
    \right)
    \right]   
}
&
    k \geq 1
\end{array}
$$
\end{definition}

\begin{definition}[Pull (non-idempotent reduction)]
\label{def:computation-pull-noni}
$$
\begin{array}{ll}
\S_\pulln^0(v) \coloneqq \bot
\ \\
\S_\pulln^1(v) \coloneqq \I(v)
\ \\
\S_\pulln^{k+1}(v) \coloneqq
\twopartdef{
    \S_\pulln^k(v)
}{
    \CPreds^k(v) = \emptyset
}{
        \E \left[ 
        \undersetp{ \, u \in \preds(v) }{\R}
        \, \P \! \left( \S_\pulln^k(u), \<u, v\> \right)
        \right]
}
&
    k \geq 1
\end{array}
$$
\end{definition}

\begin{tabular}{ll}

\begin{minipage}{\textwidth/2}

\begin{definition}[Push (idempotent reduction)]
\label{def:computation-push}
$$
\begin{array}{ll}
\S_\pushi^0(v) \coloneqq \bot
\ \\
\S_\pushi^1(v) \coloneqq \I(v)
\ \\
\S_\pushi^{k+1}(v) \coloneqq 
\E(S_n),
\ \ \ 
k \geq 1
\mbox{\ \ \ where \ }
\ \\ \t
\llet \{u_{0}, .., u_{n-1} \} \coloneqq
\CPreds^k(v) \lin
\ \\ \t
S_0 \coloneqq \S_\pushi^{k}(v)
\ \\ \t
S_{i+1} \coloneqq \R \left( S_{i}, \ \P \! \left( \S_\pushi^k(u_i), \<u_i, v\> \right) \right)
\end{array}
$$
\end{definition}

\end{minipage}

&

\begin{minipage}{\textwidth/2}

\begin{definition}[Push (non-idempotent reduction)]
\label{def:computation-push-noni}
$$
\begin{array}{ll}
\S_\pushn^0(v) \coloneqq \bot
\ \\
\S_\pushn^1(v) \coloneqq \I(v)
\ \\
\S_\pushn^{k+1}(v) \coloneqq 
\E(S_n),
\ \ \ 
k \geq 1
\mbox{\ \ \ where \ }
\ \\ \t
\llet \{u_{0}, .., u_{n-1} \} \coloneqq
\preds(v) \lin
\ \\ \t
S_0 \coloneqq \bot
\ \\ \t
S_{i+1} \coloneqq \R \left( S_{i}, \ \P \! \left( \S_\pushn^k(u_i), \<u_i, v\> \right) \right)
\end{array}
$$
\end{definition}

\end{minipage}
\end{tabular}

\caption{
    Four Iterative Reduction Methods.
    ($\name$ also incorporates another variant of Push, Non-idempotent Reduction (\apprefp{\secPushIterCorrectness})). $\CPreds^k(v)$: The predecessors of the vertex $v$ that changed in the iteration $k$
}
\label{fig:pull-push-iter}
\end{figure*}

%
\textbf{Push Model. \ }
In the pull model above,
each vertex itself pulled values from its predecessors.
In contrast, 
in the push model,
the predecessors push values to the vertex
when they are updated.
We consider the push model for idempotent and non-idempotent reduction functions in turn.

\textit{Push model with idempotent reduction ($\pushi$). \ }
The push model for idempotent reduction is represented in 
\autoref{fig:pull-push-iter},
\autoref{def:computation-push}.
The value of the vertex $v$ in the iteration $k$ is represented with $\S_\pushi^k(v)$.
In the beginning $k = 0$,
the vertices have no value $\bot$.
In the first iteration $k = 1$, 
they are initialized by the initialization function $\I$.
In subsequent iterations $k+1, \ k \geq 1$,
for each vertex $v$,
the predecessors 
$\{ u_0, .., u_{n-1} \}$
that have been changed in the previous iteration 
independently propagate their values
and reduce it with the current value of $v$.
Since the reduction function is commutative and associative,
the predecessors can apply their updates in any order. 
In each iteration,
the initial value $S_0$ of $v$ is its value in the previous iteration $k$.
For each changed predecessor $u_i$,
the propagation function $\P$ is applied
to the value of $u_i$ (from the previous iteration $k$)
and the connecting edge $\<u_i, v\>$.
The result is then reduced with the current value $S_i$ of $v$
to calculate its new value $S_{i+1}$.
Propagation and reduction by the last changed predecessor $u_{n-1}$
results in 
the value $S_n$.
The final value of $v$ in the iteration
is the application of the epilogue $\E$ to $S_n$.

\textit{Push model with non-idempotent reduction ($\pushn$). \ }
This model works
for non-idempotent (in addition to idempotent) reduction functions.
The iteration model
is represented in
\autoref{fig:pull-push-iter},
\autoref{def:computation-push-noni}.
Let the value of the vertex $v$ in the iteration $k$ be represented as $\S_\pushn^k(v)$.
%
%
Since the reduction function may not be idempotent,
in contrast to the previous model,
vertices start from the none value $\bot$
and
all the predecessors $u_i$ propagate their values
in each iteration.
For each predecessor $u_i$,
the propagate function $\P$ is applied to
the latest value $\S_\pushn^{k}(u_i)$ of $u_i$
and the edge $\<u_i, v\>$.
The resulting value is reduced with the current value of $v$.
We note that this variant 
makes all vertices active during an iteration; $\name$ also incorporates another variant (\apprefp{\secPushIterCorrectness}) where only the vertices whose values change are active and propagate their values. 
In this variant, every active predecessor $u_i$ first rollbacks its previous update before applying its new update.



The iteration models that we saw here are \emph{synchronous}.
In the synchronous model, 
vertices store the previous in addition to the new value
to propagate the previous value.
In the \emph{asynchronous} model, however, each vertex stores one value,
and vertices can propagate intermediate values.
We present four asynchronous models and their correctness in 
\appref{\secAsynchronous}.
Further, 
we present streaming iteration models and their correctness in 
\appref{\secStreamingGraphs}.

\mclearpage
\section{Specification and Fusion}
\label{sec:spec}

We 
define the core specification language and
the semantics-preserving fusion transformations:
%


\begin{wrapfigure}[30]{R}{0.6\textwidth}

\vspace{1ex}
\small
$
\begin{array}{rcll}
%

    r
        & \coloneqq &

         \underset{\Vertices}{\R} \ m
         \ \ | \ \
         r \oplus r
         \ \ | \ \
         x
         \ \ | \ \
         &
         \!\!\!\!\!\mbox{Vertex-based Red.} 
    \\

        &&
        \ilet X \coloneqq M \lin
        &

        \\ &&
        \mlet X \coloneqq E \lin
        \\ &&
        \rlet X \coloneqq R \lin
        e
%
    \\

    m
        & \coloneqq &
        

         \underset{p \in P}{\R} \F(p)
         \ \ | \ \ 
         m \oplus m
         \ \ | \ \ 
        &
        \mbox{Path-based Red.}
        \\

    &&
        \ilet X \coloneqq M
        \lin e
        \ \ | \ \ 
        x
    &
    \\

    P
        & \coloneqq &
        

        \Paths
        \ \ | \ \
        \underset{p \in P}{\mathsf{args}\R} \, \, \F(p)
        &
        \mbox{Paths}
    \\


%

%

    \R
        & \coloneqq &
            \min
            \ \ | \ \
            \max
            \ \ | \ \
            \lor
            \ \ | \ \
            \land
            \ \ | \ \
            \sum
        &
            \mbox{Reduction Fun.}
    \\

    \F
        & \coloneqq &
            \length
            \ \ | \ \
            \weight
            \ \ | \ \
            \capacity
        &
            \mbox{Path Fun.}
        \\

%
%

    \oplus    
        & \coloneqq &
            \min
            \ \ | \ \
            \max
            \ \ | \ \
            \land
            \ \ | \ \
            \lor
            \ \ | \ \
            +
            \ \ | \ \
        &
        \mbox{Operation}
        \\

        & &
            -
            \ \ | \ \
            \times
            \ \ | \ \
            \ / \             
            \ \ | \ \
            =
            \ \ | \ \
            <
            \ \ | \ \
            >
        \\




    p
        &  &
        &
        \mbox{Path Variable}
        \\

    x
        &  &
        &
        \mbox{Variable}
        \\

    e
        & \coloneqq &
            e \oplus e
            \ \ | \ \
            x
        &
        \\

    X
        & \coloneqq &
            \<X, X\>
            \ \ | \ \
            x
        &
        \\

    M
        & \coloneqq &
            \<M, M\>
            \ \ | \ \
            \R \ \F
        &
        \\

    R
        & \coloneqq &
            \<R, R\>
            \ \ | \ \
            \R \ \<\overline{x}\>
            \ \ | \ \ 
            \R \ \<\overline{d}\>
        &
        \\

    E
        & \coloneqq &
            \<E, E\>
            \ \ | \ \
            e
        &
        \\

    \CR
        & \coloneqq &
        [\ ]
         \ \ | \ \ 
         \CR \oplus r
         \ \ | \ \ 
         r \oplus \CR
        &
        \mbox{Context for $r$}
        \\

    \CM
        & \coloneqq &
        [\ ]
         \ \ | \ \ 
         \underset{\Vertices}{\R} \ \CM
        \ \ | \ \ 
        \CM \oplus m
        \ \ | \ \ 
        m \oplus \CM
        &
        \mbox{Context for $m$}
        \\

    \CMs
        & \coloneqq &
        [\ ]
        \ \ | \ \ 
        \<\CMs, M\>
        \ \ | \ \ 
        \<M, \CMs\>
        \ \ | \ \ 
        &
        \mbox{Context for $M$}
        \\
    
    &&
        \ilet X \coloneqq \CMs \lin e
        \ \ | \ \
        \\

    &&
    \multicolumn{2}{l}{
        \ilet X \coloneqq \CMs \lin 
        \ 
        \mlet X \coloneqq E \lin
        \ 
        \rlet X \coloneqq R \lin
        \ 
        e
        }
        \\

    \CRs
        & \coloneqq &
        [\ ]
        \ \ | \ \ 
        \<\CRs, R\>
        \ \ | \ \ 
        \<R, \CRs\>
        \ \ | \ \ 
        &
        \mbox{Context for $R$}
        \\

    &&
    \multicolumn{2}{l}{
        \ilet X \coloneqq M \lin 
        \ 
        \mlet X \coloneqq E \lin
        \ 
        \rlet X \coloneqq \CRs \lin
        \ 
        e
        }
        \\

    \n
        &  &
        &
        \mbox{Value}
        \\


    \v
        &  &
        &
        \mbox{Vertex Value}
        \\

        \multicolumn{3}{l}{d \colon \D_m \ \ \coloneqq \ \ \left(\Vertices(g) \mapsto \Nat\right) \union \{\bot\}}
        &
        \mbox{Sem. Dom. of $m$}
        \\

        \multicolumn{3}{l}{\n_\bot \colon \D_r \ \ \coloneqq \ \ \Nat \union \{\bot\}}
        &
        \mbox{Sem. Dom. of $r$}
        \\


\end{array}
$


\caption{
    Core specification language
}
\label{fig:spec-lang}
\end{wrapfigure}

\subsection{Core Specification Language}
\label{sec:lang-spec}

To present the crux of the fusion transformations,
we define a core specification language in \autoref{fig:spec-lang}.
%
%
It features both reduction over paths and 
reduction over vertices.
%
A computation can be specified as
a reduction $r$ over the values of \emph{vertices}.
The value of vertices can be specified as
a nested reduction $m$
over the \emph{paths} to each vertex.
%
More elaborate computations 
can be specified by 
nested path-based computations
and
applying operations between multiple vertex-based computations.
We will visit each term type in turn.

\textit{Vertex-based and path-based reductions. }
A vertex-based reduction 
$\underset{\Vertices}{\R} \ m$ 
applies a reduction function $\R$ 
to the result of path-based reductions $m$
over all vertices $\Vertices$.
The function $\R$
is a commutative and associative function
such as $\min$, 
and $\sum$.
Larger vertex-based reductions $r \oplus r'$
can be constructed using the operators $\oplus$.
A path-based reduction 
$\underset{p \in P}{\R} \F(p)$
applies a reduction function $\R$
to the result of the function $\F$ on the paths $P$.
Similar to vertex-based reductions,
larger path-based reductions $m \oplus m'$
can be constructed using 
the operators $\oplus$.
The path function $\F$ is 
the $\length$, $\weight$, or $\capacity$
of the path.
The set of paths $P$ can be either 
$\Paths$ that denotes all the paths to each vertex,
or
the restricted paths
$\underset{p \in P}{\mathsf{args}\,\R} \, \, \F(p)$ where $\R \in \{\min, \max\}$
that denotes the paths in $P$ whose $\F$ value is the minimum or maximum.
Restricted paths
lead to 
nested path-based computations.



\textit{Let forms. \ }
The fusion transformations use let terms to factor reductions.
Factored reductions are conducive to fusion.
As shown in \autoref{fig:spec-lang},
the terms $m$ and $r$ both have let forms.
The 
$m$ term constructor $\ilet X \coloneqq M \lin e$
binds 
variables $X$ to 
path-based reductions $M$
for the expression $e$.
%
The expression $e$ can apply operators $\oplus$ to the variables $X$.
Both the variables $X$ and reductions $M$ can be inductively constructed as pairs.
A single path-based reduction $M$ is simply represented as $\R \, \F$
where 
$\R$ is the reduction function 
and 
$\F$ is the path function.
Similarly, the triple-let $r$ 
constructor
$\ilet X \coloneqq M \lin
\mlet X' \coloneqq E \lin
\rlet X'' \coloneqq R \lin
e$
binds 
variables $X$ to path-based reductions $M$,
variables $X'$ to expressions $E$ (on $X$),
and
variables $X''$ to vertex-based reductions $R$ (on $X'$).
A triple-let term represents an $r$ term as
path-based reductions,
then mappings on the results,
and finally vertex-based reductions 
on the results.
We will see that 
this form enables fusion (\autoref{sec:fusion}) and
can be directly implemented (\autoref{sec:frameworks}).
%
Similar to $M$, 
the vertex-based reductions $R$ can be inductively constructed as pairs.
A single vertex-based reduction $R$ is 
$\R \ \<\overline{x}\>$ that is a reduction over tuples of vertex variables $\<\overline{x}\>$
(or 
$\R \ \<\overline{d}\>$
after the variables are substituted with map values $d$ from vertices to values).
%
To concisely represent the fusion rules,
we define the context $\CR$ to abstract the surrounding term
where a term $r$
appears.
Similarly,
we define the contexts $\CM$, $\CMs$, and $\CRs$
for the terms $m$, $M$ and $R$.


\textit{Semantics and Compositionality. }
In \appref{\secSemanticsApp},
we defined a denotational semantics for the language presented in \autoref{fig:spec-lang}. 
It defines the semantics $\sem{}$ of each term type. 
The domain $\D_m$ of a path-based computation $m$ 
on a graph $g$ is a finite map
from each vertex of $g$ to natural numbers
$\Vertices(g) \mapsto \Nat$ 
and
$\bot$ (for undefined computation).
The domain $D_r$ of a vertex-based computation $r$ 
is
the natural numbers $\Nat$ 
and 
$\bot$ (for undefined computation).
%
We prove that the semantics is compositional.
If two terms are semantically equivalent,
replacing one with the other in any context is semantics-preserving.
Compositionality of the semantics is used to prove that
the fusion transformations are semantic-preserving.
The following theorem states 
compositionality for $r$.
(The proofs and other lemmas are available in 
\appref{\secCompProof}.)
\begin{lemma}[Compositionality]
\label{lem:comp}
For all $r$, $r'$ and $\CR$,
if $\sem{r} = \sem{r'}$ then $\sem{\CR[r]} = \sem{\CR[r']}$.
\end{lemma}


\subsection{Fusion}
\label{sec:fusion}

We now present the fusion transformations.
Fusion reduces computation time by combining
separate reductions into a single reduction.
%
The transformations have three main forms: 
fusion of nested path-based reductions,
fusion of pairs of path-based reductions,
and
fusion of pairs of vertex-based reductions.
%
The result of fusion is an equivalent specification in the
triple-let form with separate terms for
path-based reduction,
mapping over vertices
and
vertex-based reduction.

The fusion rules are presented in \autoref{fig:fusion-rules}.
The top-level fusion relation $\Rightarrow_r$ is called 
$r$-fusion and transforms 
an $r$ term to another.
The other fusion relations
$\Rightarrow_m$,
$\Rightarrow_M$,
and
$\Rightarrow_R$
are called $m$-fusion, $M$-fusion and $R$-fusion,
and
transform $m$, $M$ and $R$ terms respectively.
The rule
\irule{FMInR} states that
$m$-fusions can be applied
to $m$ terms that appear in the context of 
$r$ terms.
(Both $\CM[m_1]$ and $\CM[m_2]$ in this rule are $r$ terms.)
%
We consider $m$-fusions first.
The rule
\irule{FMInM}
states that 
$m$-fusions can be applied
to $m$ terms 
in the context of
other $m$ terms.

\textit{Fusing nested path-based reductions. }
The rule \irule{FPNest} $m$-fuses nested path-based reductions to flat reductions.
Consider the nested path-based reduction
$\underset{p' \, \in \, P'}{\R} \F(p)$
where the set of paths $P'$ is 
another path-based reduction
$\underset{p \in P}{\mathsf{args}\,\R'}\, \, \F'(p)$
where $\R'$ is $\min$ or $\max$.
Let us assume that $\R'$ is $\min$.
A straightforward calculation
computes $\F'$ on the paths $P$ and finds the subset of paths $P'$ with the minimum value,
and then 
computes $\F$ on the paths $P'$ and reduces them by $\R$.
An optimized calculation
can compute both $\F'$ and $\F$ on the paths $P$ simultaneously
and only consider the pairs with the minimum first element 
to calculate the reduction $\R$ over the second elements.
To calculate the values of the path functions $\F$ and $\F'$,
this approach enumerates paths only once instead of twice.
%
(We will see in \autoref{sec:iter-reduction} 
that the calculation can further avoid 
the explicit enumeration of paths.)
Therefore, the two reductions can be fused into one reduction as
$\ilet \<x, x'\> \coloneqq \underset{p' \in P}{\R''} \, \F''(p') \lin x'$.
The new path function
$\F''(p')$ returns the pair of values $\F'(p')$ and $\F(p')$.
The new reduction function 
$\R''$ considers the first element of the two input pairs
and 
if the first element of one input is (strictly) smaller than the other, 
that input is returned.
That input takes over
because the set of paths 
for the reduction $\R$
are only those with the minimum value for $\F'$.
%
On the other hand, if the first elements of the inputs are equal, 
their second elements are reduced by $\R$
to make the second element of the output pair.
The rule \irule{FPNest} can be repeatedly applied to a deeply nested path-based reduction to flatten it to a reduction over the basic paths term $\Paths$.

\textit{Factoring, pairing and fusing path-based reductions. }
%
The rule \irule{FPRed} factors out a flat reduction to an equivalent let form.
%
%
The rule \irule{FILetBin}
fuses an operation between two let terms
to a single let term.
It pairs 
the factored reductions $M_1$ and $M_2$
of the two 
let 
terms
to keep the reductions of the resulting 
term factored.
The condition of the rule prevents
the $\free$ variables of the expression of one term from
clashing with the bound variables of another.
The rule \irule{FMInILet}
allows the factored reductions $M$ in the context of a let term
to be fused.
The rule \irule{FMPair} $M$-fuses 
a pair of factored reductions
$\left\<
\R \ \F,
\R' \ \F'
\right\>$
to a single reduction
$\R'' \ \F''$
that calculates the two reductions simultaneously.
The path function $\F''$ returns the pair of the results of $\F$ and $\F'$.
Similarly, the reduction function $\R''$ returns a pair:
the reduction of the first elements by $\R$ and the second elements by $\R'$.

\textit{Factoring into, pairing and fusing triple-let terms. }
The rules above can factor all 
path-based reductions $m$ to the let form 
and fuse factored reductions to a single one.
The next rule \irule{FVRed} 
expects path-based reductions to be in the let form.
It transforms vertex-based reductions that are applied to
path-based reductions
to an equivalent triple-let form.
The triple-let form factors 
both path-based 
and 
vertex-based reductions in separate let parts: the first and third lets respectively.
The rule \irule{FLetsBin}
fuses an operation between two triple-let terms
to a single triple-let term.
It pairs 
the factored 
path-based reductions $M$,
expression $E$,
and
vertex-based reduction $R$
of the two terms.
The rule \irule{FMInLets}
allows the factored path-based reductions $M$ in the context of a triple-let term
to be $M$-fused.
(As we saw above, 
the rule \irule{FMPair} presents $M$-fusions.)
The rule \irule{FRinLets}
allows the factored vertex-based reductions $R$ in the context of a triple-let term
to be $R$-fused.

\textit{Fusing vertex-based reductions. }
The rule \irule{FRPair} presents $R$-fusions.
It fuses a pair of $R$ reductions
$\left\< \R_1 \ x_1, \R_2 \ x_2 \right\>$
to a single $R$ reduction
$\R_3 \ \left\<x_1, x_2\right\>$.
%
Given two pairs,
the reduction function $\R_3$
returns a pair:
the reduction of the first elements by $\R_1$ and the second elements by $\R_2$.

The fusion presented above is semantic-preserving.
Terms are only fused into other terms with the same semantics.
The following theorem sates the semantics-preservation property of fusion. 
(The proofs are available in 
\appref{\secFutSoundness}.)

\begin{theorem}[Semantics-preserving Fusion]
\label{thm:sem-pres-r-main}
For all $r_1$ and $r_2$, 
if 
$r_1 \Rightarrow_r r_2$
then
$\sem{r_1} = \sem{r_2}$.
\end{theorem}

\subsection{Extensions}
\label{sec:extensions}

We now consider extensions to the core syntax and the fusion rules.
%

\textit{Common Operation Elimination. \ }
Fusion factors the path-based reduction, vertex-based mappings and vertex-based reductions 
in the triple-let form.
This form facilitates common operation elimination.
For example, if a path-based reduction is calculated twice and assigned to two sets of variables, the extra calculation can be eliminated and the result of one calculation can be assigned to both sets of variables.
The elimination rules are available in
\appref{\secComSubexpElim}.

\textit{Domain. \ }
The scalar semantic domain of the core language was confined to the natural numbers.
The domain can be simply extended to booleans, vertex identifiers
and also sets of values.
The reduction operations are extended with union $\union$ and intersection $\intersect$
and
the path functions are extended with $\head$ and $\penultimate$.
The function $\head$ returns the identifier of the head vertex of the path and 
the function $\penultimate$ returns the identifier of the penultimate (that is the vertex before  the last) of the path.
\textit{Unary operations and Literals. \ }
The path-based reductions $m$ and vertex-based reductions $r$ can be simply extended with unary operations and literals. 
Their supporting fusion rules are available in 
\appref{\secExtAppUnaryLiteral}.

\textit{Vertex Variables. \ }
%
%
We extend the core syntax with 
path terms $\Paths(v, v')$ and  $\Paths(v)$ that can specify vertex variables as source and destination.
The term $\Paths(v, v')$ specifies the set of paths from the source $v$ to the destination $v'$,
and
the term $\Paths(v)$ specifies the set of paths from any source to the destination $v$.
Thus, the source $s$ of a path-based reduction can be either a vertex $v$ or none $\bot$.
A 
factored path-based reduction
$\underset{c}{\R} \ \F$
carries its configuration $c$.
The configuration $c$ 
is a source,
or a pair of other configurations.
We also extend the syntax with
vertex-based reductions $\underset{v \in \Vertices}{\R} m$ that can bind the vertex variable $v$.
(We define this syntax extension and its fusion rules in 
\appref{\secVertexVarApp}.)

\textit{Syntactic Sugar. \ }
Syntactic sugar enable concise specifications.
For example,
the term
\linebreak
$\F (\underset{p \in P}{\mathsf{arg}\,\R} \,\, \F'(p))$
where $\R$ is either $\min$ or $\max$
first finds a path $p$ in $P$ with the minimum or maximum value for the function $\F'$
and then returns the result of applying $\F$ to $p$.
It is used to specify the \irule{BFS} use-case.
The rule \irule{FMRed} expands this term to
a path-based reduction 
in the let form
$\ilet \<x, x'\> \coloneqq \underset{p \in P}{\R'} \, \F''(p) \lin x'$.
The path function $\F''$ returns the pair of the results of $\F'$ and $\F$.
The reduction function $\R'$ returns 
the input pair with the minimum or maximum first element.
\begin{equation}
\small
{\begin{array}[t]{l}
\rulelabelp{FMRed}{FMRed}
\inferrule[FMRed] {}{
\F \, (\underset{p \in P}{\mathsf{arg}\,\R} \,\, \F'(p))
\ \ \ \    \coloneqq   \ \ \ \
\ilet \<x, x'\> \coloneqq \underset{p \in P}{\R'} \, \F''(p) \lin x'
}
\mbox{\ \ \ \ where \ \ \ \ }
\R \in \{\min, \max\}
\\
\F'' \ \coloneqq \ 
\lambda p. \ \<\F'(p), \F(p)\>
\ \ \ \ \ \ \ \ \ \ 
\R' ( \<a, b\>, \<a', b'\>) \ \coloneqq \ 
\lif \ (\R (a, a') = a) \lthen \<a, b\> \ 
\lelse \ \<a', b'\>
\end{array}
}
\end{equation}
We discuss the syntactic sugar for
(1) cardinality $|P|$,
(2) 
vertex-based reduction over a given subset of vertices
$\underset{v \in \{v_1, .., v_n\}}{\R} m$
and
(3)
vertex-based reduction constrained by a path-based reduction
$\underset{v \in \Vertices \, \land \, m'}{\R} m$
(that are used to specify the use-cases
\irule{NSP},
\irule{Radius},
and
\irule{DS})
in \autoref{fig:benchs-spec}.

\textit{Nested Triple-lets. \ }
The core syntax supports expressions that can be fused to 
a single iteration-map-reduce triple-let term.
%
We extend the core syntax 
to support nested vertex-based reductions,
and extend the fusion rules 
to fuse nested reductions.
For example, 
the use-case $\RDS$ that we saw in \autoref{fig:benchs-spec}
uses the vertex-based reduction $\Radius$ as a nested term.
%
%
Nested triple-let terms
can be
translated to a sequence of
iteration-map-reduce
rounds 
on the graph.
(In \appref{\secMultipleRounds},
we define the extension and 
show that $\RDS$ is fused to two rounds of iteration-map-reduce.)

We saw an example fusion in \autoref{fig:radius-fusion}.
More examples are available in
\appref{\secExampleFusions}.

\mclearpage
\section{Mapping Specification to Iteration-Map-Reduce}
\label{sec:mapping}

%
In the previous section, we saw a fusion process that 
transforms specifications to the following triple-let form.
As we saw 
in the final term 
in \autoref{fig:radius-fusion},
the fusion results in
the triple-let form 
shown in \autoref{fig:triplelet}.
It has
three separate operations:
the path-based reduction,
mapping expressions over their results,
and the subsequent vertex-based reduction.
%
\begin{wrapfigure}{R}{0.25\textwidth}
$
\begin{array}{l}
\footnotesize
\ilet X \coloneqq \underset{c}{\R} \ \F \lin 
\\ 
\mlet X' \coloneqq E \lin
\\ 
\rlet X'' \coloneqq \R' \ \<\overline{x'}\> \lin
e
\end{array}
$
\caption{Triple-let form}
\label{fig:triplelet}
\end{wrapfigure}
%
The three let parts can be directly mapped to three computation primitives: 
iteration, map and reduce.
Each vertex stores the variables $X$ and $X'$.
The first let is mapped to an iterative calculation 
for the path-based reduction
that results in values for the variables $X$ in each vertex.
The second let is mapped to a map operation over vertices:
given the values of the variables $X$ in each vertex,
the map operation
calculates the values of the expressions $E$
and 
stores the results in the variables $X'$ for the vertex.
%
The third let is mapped to a reduction operation over vertices:
given the values of the variables $\overline{x'}$ in $X'$ in each vertex,
the reduction operation reduces the values of $\<\overline{x'}\>$ for all vertices
and 
stores the results in the global variables $X''$.
Finally, the expression $e$ is calculated based on the values of $X''$.

\begin{figure*}
\small

%
\resizebox{.95\textwidth}{!}{
\begin{mathpar}
%
%
{
\begin{array}{l}
\rulelabelp{FMInR}{FMInR}
\inferrule[FMInR] {
m_1
\ \ \ \    \Rightarrow_m   \ \ \ \ 
m_2
}{
\CM[m_1]
\ \ \ \    \Rightarrow_r   \ \ \ \ 
\CM[m_2]
}
\\
\\
\rulelabelp{FMInM}{FMInM}
\inferrule[FMInM] {
m_1
\ \ \ \    \Rightarrow_m   \ \ \ \ 
m_2
}{
\CM[m_1]
\ \ \ \    \Rightarrow_m   \ \ \ \ 
\CM[m_2]
}
%
\end{array}
}

\begin{array}{l}
\rulelabelp{FPNest}{FPNest}
\inferrule[FPNest] {}{
\underset{p' \, \in \, \underset{p \in P}{\mathsf{args}\,\R'} \, \, \F'(p)}{\R} \F(p')
\ \ \ \    \Rightarrow_m   \ \ \ \
%
%
\ilet \<x, x'\> \coloneqq \underset{p \in P}{\R''} \, \F''(p) \lin x'
}
\\
\mbox{where \ \ }
\R' \in \{\min, \max\}
\\ \phantom{\mbox{where \ \ }}
f'' \coloneqq
\lambda p. \ \<\F'(p), \F(p)\>
\\ \phantom{\mbox{where \ \ }}
\R'' ( \<a, b\>, \<a', b'\>) \coloneqq
\\ \phantom{\mbox{where \ \ }} \t
\lif \ (a = a') \lthen \<a, \R(b, b')\> \ 
\\ \phantom{\mbox{where \ \ }} \t
\lelse \ \lif \ (\R'(a, a') = a) \lthen \<a, b\> \ 
\lelse \ \<a', b'\>
\end{array}

\rulelabelp{FPRed}{FPRed}
\inferrule[FPRed] {}{
\underset{p \in \Paths}{\R} \F(p)
\\\\
\ \ \ \    \Rightarrow_m   \ \ \ \
\\\\
\ilet x \coloneqq
\R \ \F
\lin x
}

{\rulelabelp{FILetBin}{FILetBin}
\inferrule[FILetBin] {}{
(\ilet 
X_1
\coloneqq
M_1
\lin e_1)
\ \oplus \ 
(\ilet 
X_2
\coloneqq
M_2
\lin e_2)
\\\\
\ \ \ \    \Rightarrow_m   \ \ \ \ 
\\\\
\ilet \<X_1, X_2\> \coloneqq
\<M_1, M_2\>
\lin e_1 \oplus e_2
}
\mbox{\ \ \ \ if }
{\begin{array}{l}
\free(e_1) \intersect X_2 = \emptyset   
\\
\free(e_2) \intersect X_1 = \emptyset
\end{array}}
}
%

\rulelabelp{FMInILet}{FMInILet}
\inferrule[FMInILet] {
M_1
\ \ \ \    \Rightarrow_M   \ \ \ \ 
M_2
}{
\ilet 
X
\coloneqq
\CMs[M_1]
\lin e
\\\\
\ \ \    \Rightarrow_m   \ \ \
\\\\
\ilet 
X
\coloneqq
\CMs[M_2]
\lin e
}
%
%
%
%
%
%

\begin{array}[t]{l}
\rulelabelp{FMPair}{FMPair}
\inferrule[FMPair] {}{
\left\<
\R \ \F,
\R' \ \F'
\right\>
\ \ \ \    \Rightarrow_M   \ \ \ \ 
\R'' \ \F''
}
\\ \mbox{where \ \ }
\F'' \coloneqq
\lambda p. \ \left\< \F(p), \F'(p) \right\>
\\ \phantom{\mbox{where \ \ }}
\R'' ( \<a, b\>, \<a', b'\> ) \coloneqq
\\ \phantom{\mbox{where \ \ }} \t
\left\<
\R (a, a'), \R' (b,  b') 
\right\>
\end{array}

\rulelabelp{FVRed}{FVRed}
\inferrule[FVRed] {}{
\underset{\Vertices}{\R} \ 
(\ilet X \coloneqq \R' \, \F \lin e)
\\\\
\ \ \ \    \Rightarrow_r   \ \ \ \ 
\\\\
\!\!{
\begin{array}[t]{l}
\ilet X \coloneqq \R' \, \F \lin 
\\
\mlet x \coloneqq e \lin
\\ 
\rlet x' \coloneqq \R \ x \lin
x'
\end{array}}
}

\vspace{-4ex}
\rulelabelp{FLetsBin}{FLetsBin}
\inferrule[FLetsBin] {}{
\left(
{\begin{array}{l}
\ilet X_1 \coloneqq M_1 \lin 
\\ 
\mlet X_1' \coloneqq E_1 \lin
\\ 
\rlet X_1'' \coloneqq R_1 \lin
\\
e_1
\end{array}}
\right)
\ \oplus \ 
\left(
{\begin{array}{l}
\ilet X_2 \coloneqq M_2 \lin 
\\ 
\mlet X_2' \coloneqq E_2 \lin
\\ 
\rlet X_2'' \coloneqq R_2 \lin
\\
e_2
\end{array}}
\right)
\ \Rightarrow_r  \ 
\left(
{\begin{array}{l}
\ilet \<X_1, X_2\> \coloneqq \<M_1, M_2\> \lin 
\\ 
\mlet \<X_1', X_2'\> \coloneqq \<E_1, E_2\> \lin
\\ 
\rlet \<X_1'', X_2''\> \coloneqq \<R_1, R_2\> \lin
\\
e_1 \oplus e_2
\end{array}}
\right) 
\mbox{\ if }
{\begin{array}{l}
\free(E_1) \intersect X_2 = 
\\
\free(E_2) \intersect X_1 = 
\\
\free(R_1) \intersect X_2' = 
\\
\free(R_2) \intersect X_1' = 
\\
\free(e_1) \intersect X_2'' = 
\\
\free(e_2) \intersect X_1'' = \emptyset
\\
\end{array}}
}
%

\rulelabelp{FMInLets}{FMInLets}
\inferrule[FMInLets] {
M_1
\ \ \ \    \Rightarrow_M   \ \ \ \ 
M_2
}{
\left(
{\begin{array}{l}
\ilet X \coloneqq \CMs[M_1] \lin 
\\ 
\mlet X' \coloneqq E \lin
\\ 
\rlet X'' \coloneqq R \lin
e
\end{array}}
\right)
\ \ \ \    \Rightarrow_r   \ \ \ \ 
\left(
{\begin{array}{l}
\ilet X \coloneqq \CMs[M_2] \lin 
\\ 
\mlet X' \coloneqq E \lin
\\ 
\rlet X'' \coloneqq R \lin
e
\end{array}}
\right)
}
%

\rulelabelp{FRinLets}{FRinLets}
\inferrule[FRinLets] {
R_1
\ \ \ \    \Rightarrow_R   \ \ \ \ 
R_2
}{
\left(
{\begin{array}{l}
\ilet X \coloneqq M \lin 
\\ 
\mlet X' \coloneqq E \lin
\\ 
\rlet X'' \coloneqq \CRs[R_1] \lin
e
\end{array}}
\right)
\ \ \ \    \Rightarrow_r   \ \ \ \ 
\left(
{\begin{array}{l}
\ilet X \coloneqq M \lin 
\\ 
\mlet X' \coloneqq E \lin
\\ 
\rlet X'' \coloneqq \CRs[R_2] \lin
e
\end{array}}
\right)
}
%

{\begin{array}[t]{l}
\rulelabelp{FRPair}{FRPair}
\inferrule[FRPair] {}{
\left\<
\R_1 \ x_1,
\R_2 \ x_2
\right\>
\ \ \ \    \Rightarrow_R   \ \ \ \ 
\R_3 \ \left\<x_1, x_2\right\>
}
%
\\
\mbox{where\ \ \ \ }
\R_3 ( \<a, b\>, \<a', b'\> ) \coloneqq
\\ \phantom{\mbox{where\ \ \ \ }} \t
\< \R_1 (a, a'), \R_2 (b,  b') \>
\end{array}}
\end{mathpar}
}

\caption{
    Fusion Rules
}
\label{fig:fusion-rules}
\end{figure*}

%
The two latter primitives, vertex-based mapping and reduction,
can be 
implemented by a traversal over vertices.
Since the mapping and the reduction both traverse the vertices, 
a simple optimization is to 
perform them in the same pass.
%
We consider how path-based reductions can be implemented.
We saw the iterative computation models in
\autoref{sec:iterative-models}.
Now, we present how they 
can be instantiated to implement
path-based reductions.
We first present
the correctness conditions of the iterative models
to calculate path-based reductions
(\autoref{sec:iter-reduction}),
and then present
the synthesis of iteration kernel functions based on the correctness conditions (\autoref{sec:synthesis}).
%
%

\mclearpage
\subsection{Iterative Path-based Reduction and its Correctness}
\label{sec:iter-reduction}

We now present the iterative calculation of path-based reductions.
%
We consider both the pull and push models
with both idempotent and non-idempotent reduction.
For each model, we present
correctness and termination conditions.

\textbf{Specification. \ }
In \autoref{fig:triplelet},
factored path-based reductions 
in the triple-let terms
have the form $\underset{c}{\R} \ \F$.
Considering a single reduction, $c$ is either 
none $\bot$
or
a source vertex $s$.
(We discuss a similar 
treatment
for general configurations in 
\appref{\secFactoredPathBasedRed}.)
The factored reduction 
for the former 
(with no source) is simply unrolled to
$\R_{ p \in \Paths(v) } \ \F(p)$
and
the latter 
(with the source $s$) is unrolled to
$\R_{ p \in \{p \, | \, p \in \Paths(v) \; \land \; \head(p) = s \} } \ \F(p)$.
We capture both of these reductions as the following general specification
where the condition $C(p)$ is 
$\True$ 
for the former 
and 
$\head(p) = s$ 
for the latter.

\begin{definition}[Specification]
\label{def:red-spec}
$
\Spec(v) = 
\R_{ p \, \in \, \{p \, | \, p \in \Paths(v) \; \land \; \C(p) \} } \ \F(p)
$
\end{definition}

%
%
The reduction function $\R$
returns $\bot$ on an empty set
and
returns the single element
on a singleton set.
We will see that the reduction function $\R$ is associative and commutative.
Thus,
the reduction on a set of values
is the result of applying $\R$ to the set in any order.

\textbf{Model Instantiation. \ }
Explicitly calculating the set of paths is prohibitively inefficient.
Instead, path-based reductions are calculated iteratively
based on the iterative models
that 
we saw in \autoref{sec:iterative-models}.
The iterative models are parametric in terms of the kernel functions
$\I$, $\P$ (and $\B$), $\R$, $\E$.
%
%
%
%
%
We will see in \autoref{sec:synthesis} that
the kernel functions $\I$, $\P$, $\B$ and $\R$ can be automatically synthesized
from the functions $\F$ and $\R$ of the given path-based reduction.
The epilogue function $\E$ is 
simply instantiated to the identity function.
In this subsection, 
we consider 
the correctness conditions 
on the kernel functions $\I$, $\P$, $\B$ and $\R$
such that the iterative models
calculate the specified path-based reduction.
The automatic synthesis is guided by these conditions.

\textbf{Correctness. \ } 
%
The iterative models calculate
the value $\S^k(v)$ of each vertex $v$
in iterations $k$
by propagating the values of its 
neighbor
vertices.
The iteration stops when the value of no vertex changes.
The values of the vertices
$\S^k(v)$
are expected to converge to the specification 
$\Spec(v)$.
%
We show the correctness in two steps.
%
%
%
%
First, we show that 
under certain conditions,
at the end of each iteration $k$,
the value
$\S^k(v)$
of each vertex $v$
is equal to the iteration specification $\Spec^k(v)$
for the iteration $k$.
The iteration specification $\Spec^k(v)$
is defined as
the result of reduction over paths of length less than $k$.
\begin{definition}~
\label{def:red-spec-k}
$
\Spec^k(v)
=
\undersetp{
    p \, \in \, \{ p \, | \, p \in \Paths(v) \; \land \; \C(p) \; \land \; \length(p) < k \}
}{\R} \ \F(p)
$
\end{definition}
Second, we show that under certain conditions, 
there is an index $k$ where $\Spec^k (v)$ and $Spec^{k +1} (v)$
are equal with each other and $\Spec(v)$ as well.
These two steps together show that
the values of vertices $\S^k(v)$ eventually converge to $\Spec(v)$.
We now consider the four variants of the iterative models.
%
The informal and formal proofs are available in
\apprefs{\secIterRedApp}{\secProofs}
respectively.

\textbf{Pull Model. \ }
%
We consider the correctness of the pull model to calculate path-based reductions.
We look at idempotent and non-idempotent reduction functions in turn.

The correctness of the pull models is
dependent on the conditions 
$\CB_1$ - $\CB_5$ and $\CB_7$ - $\CB_{10}$
that are presented in \autoref{fig:correct-conds};
we explain each condition in turn.
The conditions $\CB_1$ and $\CB_2$ 
require the correctness of initialization in the first iteration.
According to $\Spec^k(v)$ (\autoref{def:red-spec-k}),
in the first iteration $k = 1$,
for each vertex $v$,
the value of only the paths 
to $v$ should be considered that
(1) have length 
$l < 1$,
that is the single path $\<v, v\>$ of zero length
and
(2)
that satisfy the path condition $\C$.
Therefore, if the path condition $\C$ holds on the path $\<v, v\>$,
the value of the initialization function $\I$ should be $\F(\<v, v\>)$;
otherwise, it should be none $\bot$.
The conditions $\CB_3$ and $\CB_5$
state the requirements for the propagation function $\P$.
The condition $\CB_3$: It simply states that 
if the value of the 
vertex
is none $\bot$,
the propagated value should be none $\bot$ as well.
The condition $\CB_4$:
We saw an illustration for $\CB_4$
in \autoref{fig:overview-b} and \autoref{fig:overview-c}.
For a path $p$,
we call the value of $\F$ on $p$,
the path value of $p$.
The condition $\CB_4$ states that
if two paths $p_1$ and $p_2$ 
end in a vertex $u$ and
there is an edge $\<u, v\>$ from $u$ to a vertex $v$, then
reducing the path values of $p_1$ and $p_2$ and then propagating the result through $\<u, v\>$
is the same as
%
reducing the path values of the two extended paths 
$p_1 \cdot \<u, v\>$
and 
$p_2 \cdot \<u, v\>$.
(The path $p \cdot e$ denotes the extension of the path $p$ at the end with the edge $e$.)
Intuitively, this condition states that 
the local reduction and propagation of the iterative models
effectively calculate 
reduction over paths.
The condition $\CB_5$:
%
Vertices that have only a single incoming path $p$
do not receive multiple values to be reduced.
%
For such vertices,
$\CB_5$ 
states that
the 
propagation of the path value of $p$ over the outgoing edge $e$ 
is equal to
the path value of the extended path $p \cdot e$.
The condition $\CB_7$ - $\CB_{10}$ state the required properties of the reduction function $\R$.
The none value $\bot$ should be the identity value of $\R$,
and
$\R$ should be commutative, associative and idempotent.
For example, 
given the factored path-based reduction
$\underset{s}{\min} \, \length$ 
for the shortest path use-case \irule{SSSP},
the correct kernel functions
that we saw in
\autoref{fig:iter-kern-exam}
satisfy the conditions above.

\textit{Pull model with idempotent reduction. \ }
%
%
The following theorem states that if the conditions above hold,
then 
the value $\S_\pulli^k(v)$ that the pull model with idempotent reduction
(\autoref{def:computation-pull})
calculates
complies with the specification $\Spec^k(v)$.
\begin{theorem}[Correctness of Pull (idempotent reduction)]
\label{thm:iteration-pull}
For all $\R$, $\F$, $\C$, $\I$, $\P$, and $k \geq 1$,
if the conditions $\CB_1$ - $\CB_{9}$ hold,
then
$
\S_\pulli^k(v) = 
\Spec^k(v)
$.
\end{theorem}

\begin{wrapfigure}[24]{R}{0.5\textwidth}

\small

$
\begin{array}{l}
\mbox{Initialization:}
\ \\ \t
\CB_1 \colon \forall v. \ 
\C(\<v, v\>) \rightarrow \I(v) = \F(\<v, v\>)
\ \\ \t
\CB_2 \colon \forall v. \ 
\lnot \C(\<v, v\>) \rightarrow \I(v) = \bot
\ \\[2pt]

\mbox{Propagation:}
\ \\ \t
\CB_3 \ (\mbox{None Propagation})\colon
\ \\ \t\t
\forall e. \ 
\P(\bot, e) = \bot
\ \\[2pt]

\t
\CB_4 \ (\mbox{Aggregate Propagation})\colon 
\\ \t\t
\forall p_1, p_2, v. \ 
\\ \t\t\t 
\tail(p_1) = \tail(p_2)
\rightarrow
\\ \t\t\t
\llet u \coloneqq \tail(p_1) \lin
\\ \t\t\t
\P\left[
    \R(\F(p_1), \F(p_2)), \ 
        \<u, v\> 
\right] = 
\\ \t\t\t
\R\left[
    \F(p_1 \cdot \<u, v\>), \ 
    \F(p_2 \cdot \<u, v\>)
\right]
\ \\ \t
\CB_5 \ (\mbox{Single Path})\colon 
\\ \t\t
\forall p, e. \ 
\P(
    \F(p), \ 
       e ) = 
\F(
    p \cdot e) \ 
    
\ \\[2pt]

\mbox{Reduction:}
\ \\ \t
\CB_6 \ (\mbox{Identity})\colon 
\ \\ \t\t
\forall n. \ \R(n, \bot) = n
\ \\ \t
\CB_7 \ (\mbox{Commutativity})\colon 
\ \\ \t\t
\forall n, n'. \ \R(n, n') = \R(n', n)
\ \\ \t
\CB_8 \ (\mbox{Associativity})\colon 
\ \\ \t\t
\forall n, n', n''. \ \R(\R(n, n'), n'') = \R(n, \R(n', n''))
\ \\ \t
\CB_{9} \ (\mbox{Idempotency})\colon 
\ \\ \t\t
\forall n. \ \R(n, n) = n
\ \\[2pt]

\mbox{Termination:}
\ \\ \t
\CB_{10} \colon
%
\forall p. \ 
\R(\F(p), \F(\simple(p))) = \F(\simple(p))
\end{array}
$

\caption{
    Correctness and termination conditions
}
\label{fig:correct-conds}
\end{wrapfigure}

\textit{Pull model with non-idempotent reduction. \ }
%
We saw 
the pull model with non-idempotent reduction
$\S_\pulli^k(v)$
in \autoref{def:computation-pull-noni}.
We show that it can correctly calculate path-based reductions with
non-idempotent
(in addition to idempotent) reduction functions.
%
%
For instance,
consider the factored path-based reduction 
$\underset{s}{\sum} \, 1$ 
that counts the number of paths from the source $s$;
the reduction function sum $\sum$ is non-idempotent.
%
The initialization function is instantiated to
$\I = \lambda v. \ 1$
and
the propagation function
is instantiated to
$\P = \lambda n, e. \ n$
that simply propagates the value of the predecessor.

The following theorem states that 
if the conditions above except idempotency hold
and
the given source vertex is not on any cycle
then
the pull model with non-idempotent reduction 
$\S_\pulln^k(v)$
complies with
the specification $\Spec^k(v)$.

\begin{theorem}[Correctness of Pull (non-idempotent reduction)]
\label{thm:iteration-pull-noni}
For all $\R$, $\F$, 
$\I$, $\P$, $k \geq 1$, and $s$,
let $\C(p) = (\head(p) = s)$,
if the conditions $\CB_1$ - $\CB_8$ hold, 
and
$s$ is not on any cycle,
$
\S_\pulln^k(v) = 
\Spec^k(v)
$.
\end{theorem}

\textbf{Push Model. \ }
We now consider the correctness of the push model to calculate path-based reductions.
%
%

\textit{Push model with idempotent reduction. \ }
The following theorem states that 
if the conditions $\CB_1$ - $\CB_{9}$ hold,
the value $\S_\pushi^k(v)$ that the push model with idempotent reduction 
(\autoref{def:computation-push})
calculates
complies with
the specification $\Spec^k(v)$.

\begin{theorem}[Correctness of push (idempotent reduction)]
\label{thm:iteration-push}
For all $\R$, $\F$, $\C$, $\I$, $\P$, and $k \geq 1$,
if the conditions $\CB_1$ - $\CB_9$ hold,
$
\S_\pushi^k(v) = 
\Spec^k(v)
$.
\end{theorem}

\textit{Push model with non-idempotent reduction. \ }
%
Similarly, the following theorem 
states the correctness of
the model with non-idempotent reduction 
(\autoref{def:computation-push-noni}).

\begin{theorem}[Correctness of Push (non-idempotent reduction)]
\label{thm:iteration-push-noni}
For all $\R$, $\F$, 
$\I$, $\P$, $k \geq 1$, and $s$,
let $\C(p) = (\head(p) = s)$,
if the conditions $\CB_1$ - $\CB_8$ hold, 
and
$s$ is not on any cycle,
$
\S_\pushn^k(v) = 
\Spec^k(v)
$
\end{theorem}



\textbf{Termination. \ }
We showed that all the four iteration models
comply with the iteration specification $\Spec^k(v)$
in every iteration $k$.
We now show that under certain conditions,
there exists an iteration $k$ where
$\Spec^k(v)$ (\autoref{def:red-spec-k}) stays unchanged
and
converges to the original specification $\Spec(v)$ (\autoref{def:red-spec}).
The observation is that iterations 
incrementally consider longer paths;
however, longer paths do not necessarily yield new information.
For example,
in the shortest path use-case \irule{SSSP},
after considering all the simple paths,
the longer paths (that are cyclic) cannot lead to
shorter paths (in graphs with non-negative edges).
Given a path $p$,
we call the path 
that results from removing its cycles 
the simplification $\simple(p)$ of $p$.
%
In the shortest path use-case \irule{SSSP},
the reduction function $\R$ is $\min$ and
the path function $\F$ is $\weight$.
Reducing 
the $\F$ value of $\simple(p)$
with 
the $\F$ value of $p$
results in the former.
In other words,
simplified paths are enough 
to arrive at the same result 
for the reduction.
%
%
We capture this property as the condition $\CB_{10}$ in \autoref{fig:correct-conds}.
The following theorem states 
that $\CB_{10}$ is sufficient for termination.

\begin{theorem}[Termination]
\label{thm:iteration-termination}
For all $\R$, $\F$, and $\C$, 
if 
the graph is acyclic or
the condition $\CB_{10}$ holds, then
there exists $k$ such that for every $k' \geq k$,
$\Spec^{k'}(v) = \Spec(v)$.
\end{theorem}

Let $l$ be the length of the longest simple path to the vertex $v$.
After the iteration $k = l+1$, the value of $\Spec^{k}(v)$
stays unchanged.
This is because the reduction over the paths of length greater than $l$ does not change the value 
of $\Spec^{k}(v)$.
Any path $p$ of length greater than $l$ is not simple, i.e., it includes a cycle.
This is refuted if the graph is acyclic.
Otherwise,
the simplification of $p$, $\simple(p)$, is already in the set of paths of length less than $l + 1$
and by the condition $\CB_{10}$, 
reducing 
the path value of $p$ with
the path value of $\simple(p)$
results in the path value of $\simple(p)$.


An immediate corollary of the above theorem is that 
if the graph is acyclic or the condition $\CB_{10}$ holds,
then 
the above iteration models eventually terminate and converge to the specification
(if the corresponding conditions in 
\autoref{thm:iteration-pull} to
\autoref{thm:iteration-push-noni} hold).
The final iteration is simply the maximum value of $k$ from the above theorem for all vertices.
%
For example, the corollary for the pull model for idempotent reduction functions 
is the following.
%
%
\begin{corollary}[Termination for pull model with idempotent reduction]
\label{cor:iteration-push}
For all $\R$, $\F$, $\C$, $\I$, and $\P$,
if the conditions $\CB_1$ - $\CB_9$ hold,
and
the graph is acyclic or the condition $\CB_{10}$ holds,
then there exists an iteration $k$ such that
$
\S_\pulli^k(v) = 
\Spec(v)
$.
\end{corollary}

\mclearpage

\mclearpage
\subsection{Synthesis of Iterative Reduction}
\label{sec:synthesis}

In this subsection, 
we use the correctness conditions
presented in the previous subsection
to automatically synthesize 
correct-by-construction 
kernel functions.


Given a path-based reduction
$\underset{c}{\R} \ \F$,
the goal is to 
synthesize the 
kernel functions
that are used by 
a target iterative reduction model.
%
For example, consider the push iteration model with idempotent reduction
that we saw in 
\autoref{def:computation-push}.
By \autoref{thm:iteration-push}, we need to find the functions $\I$, $\P'$ and $\R'$  
such that 
the conditions $\CB_1$ - $\CB_{10}$ (presented in \autoref{fig:correct-conds}) hold.
We use these conditions to synthesize the functions $\I$, $\P'$ and $\R'$.
In particular, 
(1) we use the initialization conditions 
$\CB_1$ - $\CB_2$ to synthesize $\I$.
(2) We use the propagation condition $\CB_4$ and $\CB_5$ to synthesize $\P$ and then wrap it in the following function $\P'$ to handle none $\bot$ values and satisfy the condition $\CB_3$.
\begin{equation}
\label{eq:option-propagate}
\nonumber
\begin{array}{l}
\P' \coloneqq \lambda n, e. \ 
\lif \ (n = \bot) \ \lreturn \ \bot \ 
\lelse \ \lreturn \ \P(n, e)
\end{array}
\end{equation}
(3) We check the conditions $\CB_7$ - $\CB_{9}$ for the reduction function $\R$.
Then, we wrap 
$\R$ in the following reduction function $\R'$ to handle none $\bot$ values and satisfy the condition $\CB_6$.
\begin{equation}
\label{eq:option-reduce}
\nonumber 
\begin{array}{l}
\R' \coloneqq \lambda a, b. \ 
\lif \ (a = \bot) \ \lreturn \ b \ 
\lelse \ \lif \ (b = \bot) \ \lreturn \ a
\\ \phantom{\R' \coloneqq \lambda a, b. \ \lif \ (a = \bot) \ \lreturn \ b \ \lelse \ }
\lelse \ \lreturn \ \R(a, b)
\end{array}
\end{equation}
If the conditions $\CB_7$ - $\CB_{9}$ hold for 
$\R$,
they hold for 
$\R'$ 
as well.


To find candidate expressions for the body of the kernel functions,
we apply a type-guided enumerative search.
%
It 
enumerates expressions 
from the grammar that we saw in 
\autoref{fig:iter-kern-exam}a
in the order of increasing size.
To support overloaded operators,
the expression constructors have union types.
To synthesize an expression of the given type,
the search only considers expression constructors that return that type.
It then recursively searches for the arguments and
uses memoization to avoid redundant enumeration.

The procedure 
that synthesizes $\P$
starts by memoizing
expressions of size one, literals and variables,
to make them available for the synthesis of the body of $\P$.
%
%
Let $T$ be the return type of $\F$;
vertices store values of type $T$.
The propagation function $\P$ takes 
a value stored at a vertex (of type of $T$)
and an edge (of type $\mathsf{Edge}$)
and returns a vertex value (of type $T$).
Thus, 
the two input parameters of
type $T$ 
and
$\mathsf{Edge}$
are 
memoized as available expressions.
%
Then,
candidate bodies for $\P$
of type $T$
of increasing sizes 
are incrementally obtained.
%
%
%
%
A candidate propagation function 
$\lambda n, l. \ e$ 
is correct if
the conditions
$\CB_4$ and
$\CB_5$
are valid
when $\P$ is replaced by the candidate.
%
%
%
The context of the validity check
$\F; \R; \Gamma$
is
the definition of the functions $\F$ and $\R$ from the given path-based reduction,
and
a set of assertions $\Gamma$
that define basic graph functions and relations.
%
%
%
%
We model paths as lists of vertices and define graph functions and relations 
including the path functions
$\mathsf{length}$, 
$\mathsf{weight}$, $\mathsf{punultimate}$ and $\mathsf{capacity}$
in 
the combination of the quantified uninterpreted functions and list theories.
More details 
including the assertions $\Gamma$ 
are available in
\appref{\secSynthesisApp}.
The synthesis of the other kernel functions is similar.

%

For termination,
we check a stronger condition than $\CB_{10}$.
Instead of removing cycles, we remove an edge:
for every path $p$ and edge $e$,
if reducing
the $\F$ value of $p$
with
the $\F$ value of $p \cdot e$
results in the former,
then the reduction is terminating.
%
(For synthesis in the push variant that requires rollback,
after the propagation function is synthesized,
a condition on the propagation $\P$ and rollback $\B$ functions is used
to synthesize the rollback function (\apprefp{\secPushIterCorrectness}).)

\mclearpage
\section{Experimental Results}
\label{sec:implementation}
\label{sec:experiments}

%

\textbf{Implementation. \ }
We implemented 
the \name \ synthesis tool in three parts:
fusion, synthesis and backends.
%
%
The fusion phase closely follows the fusion rules (of \autoref{sec:fusion})
using on the visitor pattern.
%
%
%
%
%
The synthesis phase uses
the Z3 SMT solver to check the validity of the correctness conditions.
%
%
%
%
\label{sec:frameworks}
$\name$ incorporates 
a dedicated backend for 
each framework.
Each backend generates a framework-specific C++ file containing the initialization $\I$, propagation $\P$, (if needed rollback $\B$) and reduction $\R$ functions. 
%
(The different mappings for each of the target frameworks are presented in \appref{\secFrameworksAll}.)
$\name$ can be modularly extended with backends for new frameworks.
\textbf{Platform and benchmarks. \ }
We performed the experiments on an 4-node cluster, each with 8 cores and 64GB memory. 
The experiments for frameworks that are exclusively for shared memory are performed on one of these nodes.
The nodes are connected via 40Gbps InfiniBand network, and they run CentOS 7.4 Linux 3.10.0.x86\_64. All programs are compiled with gcc-5.1.0 (for Ligra, GridGraph and GraphIt) and mpich-3.2.1 (for PowerGraph and Gemini).
%
%
\autoref{table:datasets} 
lists the characteristics of the input datasets.
We executed each experiment 5 times and reported the average.

\begin{wrapfigure}[7]{R}{0.41\textwidth}
\footnotesize
\centering
 \begin{tabular}{l c c c} 
 \hline
 Dataset & \textbar V\textbar& \textbar E\textbar & Data \\ [0.5ex] 
 \hline
 LiveJournal (LJ) & 4.8M & 68.9M & 1.1G \\ 
 twitter-www (TW) & 41.6M & 1.4B & 23G \\
 twitter-mpi (TM) & 52.5M & 1.9B & 28G \\
 Friendster (FR) & 65.6M & 1.8B & 31G \\ [1ex] 
 \hline
 \end{tabular}
 \caption{\label{table:datasets}Dataset Characterization}
\end{wrapfigure}

\textbf{Evaluation Summary. \ }
To evaluate the $\name$ synthesis tool,
we compare the synthesized code with available handwritten versions in the frameworks.
Experimental results show that the synthesized code either matches or outperforms handwritten code.
We also study the effect of fusion on the performance of the generated code. 
Experimental results show that fusion can lead up to 
4$\times$ 
faster execution time compared to the 
unfused baseline.
%
%
We also report the synthesis time.
$\name$ can 
efficiently generate programs in 
less than a minute.


\newcommand{\cbox}[1]{\raisebox{\depth}{\fcolorbox{black}{#1}{\null}}}
\begin{figure*}

\begin{subfigure}{.1\textwidth}
\begin{flushleft}
\small
\cbox{amethyst} LJ
\\
\cbox{aureolin} TW
\\ 
\cbox{persianred} TM 
\\
\cbox{parisgreen} FR
\end{flushleft}
\end{subfigure}
\begin{subfigure}{.89\textwidth}
\small
\centering
\includegraphics[scale=.89]{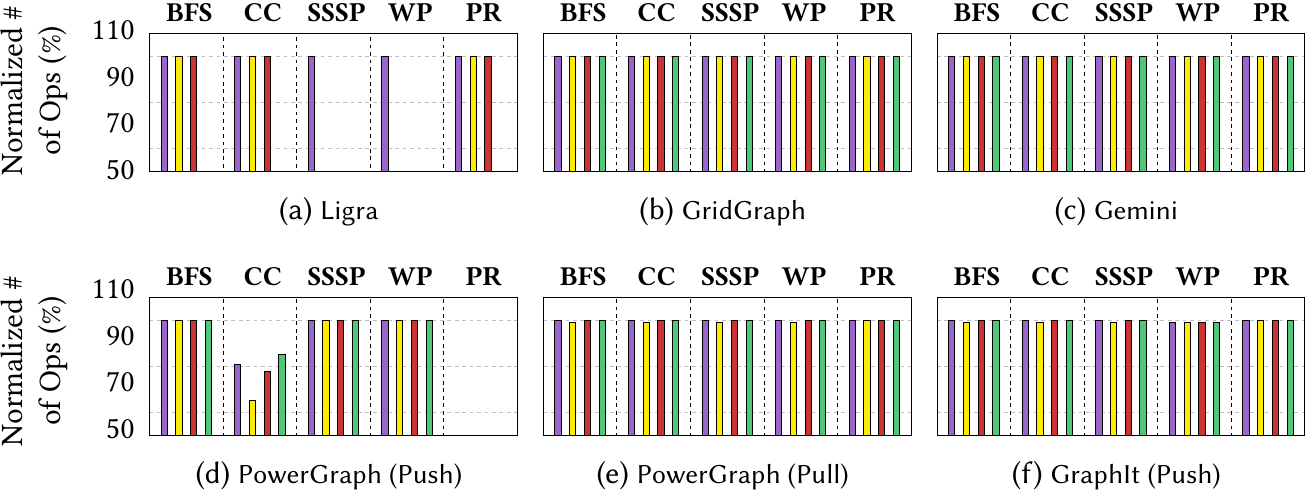}
\end{subfigure}
\caption{Edge-work ratio: the normalized number of edges processed by the synthesized programs over the handwritten programs. 
The target for the five basic use-cases is 100\%; lower ratios show further improvement.
Missing bars are due to either missing handwritten use-cases (\textsf{PR}) or 
a 
timed-out
after 24 hours.
}
\label{fig:edge-works-nonfusion}
\end{figure*}

\captionsetup{skip=0pt}
\renewcommand{\arraystretch}{.9}
\setlength{\tabcolsep}{3pt}
\begin{table}
\centering
\scriptsize
\caption{\label{table:exectime-simple}Execution times (in seconds). H: Handwritten, S: Synthesized, R: the ratio $\frac{H}{S}$. Missing cells are due to either missing handwritten use-cases (\textsf{PR}) or the corresponding experiment timed-out
after 24 hours.
}
\resizebox{\textwidth}{!}{
\begin{tabular}{c | c | c | c | c | c | c | c | c | c | c | c | c | c | c | c | c | c | c | c }
\toprule
\centering
\multirow{2}{*}{Prog.} & \multirow{2}{*}{Input} & \multicolumn{3}{c|}{Ligra} & \multicolumn{3}{c|}{GridGraph}& \multicolumn{3}{c|}{Gemini} & \multicolumn{3}{c|}{PowerGraph (Push)} & \multicolumn{3}{c|}{PowerGraph (Pull)}& \multicolumn{3}{c}{GraphIt (Push)}\\
\cline{3-20}\\[-2mm]
 & & H & S & \textbf{R} & H & S & \textbf{R} & H & S & \textbf{R} & H & S & \textbf{R} & H & S & \textbf{R} & H & S & \textbf{R}\\
 \cline{1-20}\\[-2mm]
\multirow{4}{*}{$\BFS$}  & LJ & 3.28 & 3.17 & \textbf{1.03} & 1.56 & 1.56 & \textbf{1} & 0.25 & 0.26 & \textbf{0.96} & 6.1 & 6.4 & \textbf{0.95} & 11.6 & 11.2 &  \textbf{1.04}& 0.39 & .35 &\textbf{1.1}\\
  &TW & 132 & 128 & \textbf{1.03} & 210 & 195 & \textbf{1.07} & 3.1 & 4.1 & \textbf{0.75} &  45.6 & 39 & \textbf{1.17} & 117.9 & 112.6 & \textbf{1.05} & 6 & 5.2 &\textbf{1.15}\\
  & TM & 260 & 243 & \textbf{1.06} & 487 & 472 & \textbf{1.03} & 55 & 46 & \textbf{1.2} & 38.6 & 35.5 & \textbf{1.09} & 93.6 & 97 & \textbf{0.96} & 12.4 & 10.6 &\textbf{1.17}\\
  & FR & - & - & - & 521 & 532 & \textbf{0.97} & 16 & 17.3 & \textbf{0.92} & 51.3 & 64.8 & \textbf{0.79} & 122.1 & 110.3& \textbf{1.11}& 61 & 71.2 &\textbf{0.86}\\
\bottomrule\\[-2mm]
\multirow{4}{*}{$\CC$}  & LJ & 1.69 & 1.78 & \textbf{0.94} & 2.21 & 2.22 & \textbf{0.99} & 0.79 & 0.81 & \textbf{1} &  13 & 11.7 & \textbf{1.11} &  22 & 19.1 & \textbf{1.15}&0.48 & 0.4&\textbf{1.15}\\
  & TW & 131 & 120 & \textbf{1.09} & 230 & 214 & \textbf{1.07} & 7.3  & 7.7 & \textbf{0.94} &  97.6 & 66.4 & \textbf{1.47} & 167.3 & 157.4 & \textbf{1.06}&10.6 &10.6 &\textbf{1}\\
  & TM & 184 & 187 & \textbf{0.98}  & 432 & 423 & \textbf{1.02} & 16 & 19.5 & \textbf{0.82} & 128.8 & 95.5 & \textbf{1.35} & 219.4 & 196.3& \textbf{1.12}& 316& 351 &\textbf{0.9}\\
  & FR & - & - & - & 606 & 599 & \textbf{1.01} & 51 & 48 & \textbf{1.06} & 223.9 & 184.3 & \textbf{1.21} & 375 & 353& \textbf{1.06}& 26.6 &22.8 &\textbf{1.16}\\
\bottomrule\\[-2mm]
\multirow{4}{*}{$\SSSP$}  & LJ & 4.61 & 4.8 & \textbf{0.96} & 2.42 & 2.1 & \textbf{1.15} & 0.38 & 0.4 & \textbf{0.95} & 6.1 & 6.6 & \textbf{0.92} & 12.6 & 12.8 & \textbf{0.98}& 0.41 & 0.47 &\textbf{0.88}\\
  & TW & - & - & - &  201 & 205 & \textbf{0.98} &  4.1 & 5  & \textbf{0.82} & 34.7 & 33.2 & \textbf{1.05} & 105.3 & 98.7 & \textbf{1.07}& 8.8 & 9.2 &\textbf{0.95}\\
  & TM & - & - & - & 490 & 487 & \textbf{1} & 10 & 9.1 & \textbf{1.09} & 32 & 34.3& \textbf{0.93} & 90.6 & 84.9& \textbf{1.07}& 115.2 & 116.8 &\textbf{0.98}\\
  & FR & - & - & - & 572 & 570 & \textbf{1} & 21 & 24 & \textbf{0.87} & 48.6 & 41.8 & \textbf{1.16} & 116 & 121.9& \textbf{0.95}& 83 & 77.8 &\textbf{1.07}\\
\bottomrule\\[-2mm]
\multirow{4}{*}{$\WP$}  & LJ & 6.53 & 6.4  & \textbf{1.02} & 3.46 & 3.2 & \textbf{1.08} & 0.43 & 0.41 & \textbf{1.04} &  5.7 & 5.8 & \textbf{0.98} & 10.7 & 10.5 & \textbf{1.02}& 0.49 & 0.44 &\textbf{1.11}\\
  & TW & - & - & - & 245 & 242  & \textbf{1.01} & 4.3 & 4.8 & \textbf{0.9} & 35 & 34.8 & \textbf{1.01} & 110.5 & 105.7& \textbf{1.05}& 9.8 & 8.8 &\textbf{1.11}\\
  & TM & - & - & - & 479 & 498 & \textbf{0.96} & 9 & 7 & \textbf{1.2} & 34.4 & 31.4 & \textbf{1.10} & 94.2 & 81.9& \textbf{1.15}& 4762 & 5212 &\textbf{0.9}\\
  & FR & - & - & - & 551 & 545 & \textbf{1.01} & 26 & 24 & \textbf{1.08} & 47 & 46.4 & \textbf{1.01} & 109.1 & 102.3& \textbf{1.07}& 169 & 165 &\textbf{1.01}\\
  \bottomrule\\[-2mm]
\multirow{4}{*}{PR}  & LJ & 132 & 120 & \textbf{1.1} & 44 & 37 & \textbf{1.1} & 21 & 21 & \textbf{1} &  - & - & \textbf{-} & 80 & 80 & \textbf{1}&11.8 &11.4 &\textbf{1.03}\\
  & TW & 8290 & 8270 & \textbf{1} & 1000 & 908  & \textbf{1.1} & 282 & 400 & \textbf{0.7} & - & - & \textbf{-} & 1128 & 1041 & \textbf{1.08}&319 &331 &\textbf{0.96}\\
  & TM & 13500 & 14700 & \textbf{0.91} & 1399 & 1441 & \textbf{0.97} & 880 & 860 & \textbf{1.02} & - & - & \textbf{-} & 1157 & 1078 & \textbf{1.07}&596 & 613&\textbf{0.97}\\
  & FR & - & - & - & 1023 & 995 & \textbf{1.02} & 590 & 577 & \textbf{1.02} & - & - & \textbf{-} & 601 &548 & \textbf{1.09}& 260& 280&\textbf{0.93}\\
\bottomrule
\end{tabular}
}
\end{table}

\textbf{Synthesized Matching Handwritten. \ }
%
We used five use-cases \irule{BFS}, \irule{CC}, \irule{SSSP},
$\WP$ (widest path)
and 
\irule{PR} (page-rank)
to compare
the performance of the synthesized programs and their equivalent handwritten programs.
We adopted the hand-written implementations of \irule{BFS}, \irule{CC},
\irule{SSSP}
and
\irule{PR}
that are available in the frameworks,
and 
developed $\WP$ based on \irule{SSSP} by changing the path function.
We ran \textsf{PR} on the input graphs until convergence.
%
To thoroughly study the performance of the synthesized programs, 
we measure two metrics: the \emph{number of edges processed} and the \emph{execution time}.
%
%
%
The \emph{number of edges processed} by a program 
indicates 
how many times propagation happens across edges, 
and hence, the amount of computation performed throughout the execution. 
%
%
Since path-based calculations have asynchronous semantics \cite{vora2017thesis}, 
vertex values can take different execution paths before converging to the final results,
resulting in different amount of edges processed for the same use-case.
This means, 
a poorly synthesized program can perform redundant edge computations 
%
but can still converge to the correct result,
and hence, 
we compare the number of edges processed
by synthesized and handwritten programs.
%
%
%
%
%
The second metric is the \emph{execution time}. 
Although the execution time is primarily
dependent on the number of processed edges,
it is also dependent on
the efficiency of the kernel functions
which 
can be optimized by
generating minimal vertex and edge variables
and
the minimal use of atomic operations. 
\captionsetup{skip=0pt}
\renewcommand{\arraystretch}{.9}
\setlength{\tabcolsep}{3pt}
\begin{table}
\renewcommand\thetable{2}
\centering
\footnotesize
\caption{\label{table:metrics}Metrics for comparing handwritten and synthesized code. H: Handwritten, S: Synthesized. 
(PowerGraph 
does not require the user to write atomic operations and hence 
the last two columns are zeros.) }
\resizebox{\textwidth}{!}{
\begin{tabular}{c | c | c | c | c | c | c | c | c | c | c | c | c | c | c | c | c | c | c | c | c }
\toprule
\multirow{3}{*}{Prog.} & \multicolumn{10}{c|}{Vertex Data Size (bytes) \ :: \ Edge Data Size (bytes)} & \multicolumn{10}{c}{\# Atomics Per Edge}\\
\cline{2-21}
 & \multicolumn{2}{c|}{Ligra} & \multicolumn{2}{c|}{GridGraph}& \multicolumn{2}{c|}{Gemini} & \multicolumn{2}{c|}{PowerGraph}&\multicolumn{2}{c|}{GraphIt} & \multicolumn{2}{c|}{Ligra} & \multicolumn{2}{c|}{GridGraph}& \multicolumn{2}{c|}{Gemini} & \multicolumn{2}{c|}{PowerGraph}&\multicolumn{2}{c}{GraphIt}\\
\cline{2-21}\\[-2.5mm]
 & H & S & H & S& H & S & H & S & H & S & H & S& H & S & H & S& H & S& H & S\\
 \cline{1-21}\\[-2.5mm]
$\BFS$  & 8::0 & 8::0 & 8::0 & 8::0 & 8::0 & 8::0 & 12::0 & 12::0 & 4::0 & 8::0 & 1 & 1 & 1 & 1 & 1 & 1 & 0 & 0 & 1 & 1\\
$\CC$  & 4::0 & 4::0 & 4::0 & 4::0 & 4::0 & 4::0 & 8::0 & 8::0 &4::0  &4::0  & 1 & 1 & 1 & 1 & 1 & 1 & 0 & 0 & 1 & 1\\
$\SSSP$  & 4::4 & 4::4 & 4::4 & 4::4 & 4::4 & 4::4 & 4::4 & 4::4 & 4::4 &4::4  & 1 & 1 & 1 & 1 & 1 & 1 & 0 & 0 & 1 & 1\\
$\WP$  & 4::4 & 4::4 & 4::4 & 4::4 & 4::4 & 4::4 & 4::4 & 4::4 & 4::4  & 4::4 & 1 & 1 & 1 & 1 & 1 & 1 & 0 & 0 & 1 & 1\\
$\textsf{PR}$  & 4::0 & 4::0 & 4::0 & 4::0 & 8::0 & 8::0 & 8::0 & 8::0 & 4::0&4::0 & 1 & 1 & 1 & 1 & 1 & 1 & 0 & 0 & 0 & 0\\
\bottomrule
\end{tabular}
}
\end{table}

\textit{Assessment. \ }
\autoref{fig:edge-works-nonfusion} shows the number of edges processed by the synthesized 
programs normalized w.r.t. that processed by the handwritten programs 
(i.e.\ the former divided by the latter),
that we call edge-work ratio. 
We observe that the synthesized programs
match or outperform
handwritten programs.
They process the same number of edges compared to 
handwritten programs in Ligra, GridGraph and Gemini. 
On PowerGraph and GraphIt, the synthesized programs 
process fewer edges.
While the reduction is less than 1\% 
in most cases, 
it is visible 
for the use-case \irule{CC} in the push model
and slightly visible 
for the TW input graph in the pull model and \textsf{WP} in the GraphIt.
This is because the
synthesized programs initialize vertices by directly mapping over them, 
while the handwritten
programs perform initialization in the \texttt{apply} step of the first iteration. 
%
Initialization in the \texttt{apply} step 
results in additional edge propagations 
in the first iteration.
This has a high impact on the edge-work ratio for \irule{CC} in the push model 
(down to 77\%).
Since all vertices need to be initialized in this use-case, 
the handwritten version
unnecessarily processes all edges
in the first iteration.

\autoref{table:exectime-simple}
shows the execution times of the handwritten programs (H), synthesized programs (S) and their relative ratio (R), i.e., former divided by the latter.
We observe that the execution time is closely related to the number of processed edges (\autoref{fig:edge-works-nonfusion}).
The performance of the handwritten and synthesized code is similar in most cases.
The synthesized \irule{CC} for PowerGraph in the push model, performs 28\% faster in average.
%

Running the same program (either synthesized or handwritten)
multiple times shows a variance in the execution time due to 
variances from the runtime environment.
To have a more precise comparison,
we further compare the number of atomic operations per edge computation,
and 
the size of state maintained per vertex and edge,
which are two major sources of inefficiency in graph computations. 
The results are shown in \autoref{table:metrics}.
The number of atomic operations per edge, the size of vertex, and edge states
are equal in 
synthesized and handwritten programs.

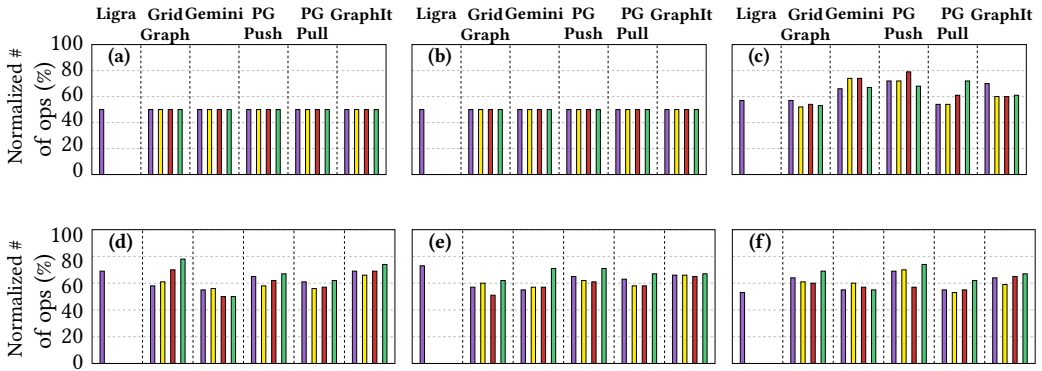
\begin{figure*}[t]
\centering
\vspace{-.3cm}
\begin{subfigure}{0.08\textwidth}  
\centering
\begin{tikzpicture}
\tikzstyle{every node}=[font=\small]
\begin{axis}[
    style=thin,
    height=3.24cm,  
    width=2cm,
    ymin=0, 
    ymax=100,
    xmin = 0,
    xmax = 1,
    ytick={0, 20, 40, 60, 80, 100},
    ylabel={\fontsize{8}{6} \selectfont Normalized \# \\ of ops (\%)},
    ylabel style={align=center,at={(1,-.82)},xshift=2.15cm},
    ytick style={draw=none},
    xtick style={draw=none},
    x tick label style={draw=none},
    axis line style={draw=none},
    hide x axis,        
    axis y line*=left,
]
\end{axis}
\end{tikzpicture}
\end{subfigure}
\hspace{-.15cm}
\begin{subfigure}{.30\textwidth}  
\centering
\resizebox{\textwidth}{!}{
\begin{tikzpicture}
\tikzstyle{every node}=[font=\Huge]
\begin{axis}[
    style=thick,
    height=7cm, 
    ymin=0, ymax=100, 
    width=14cm,
    xmin = 0,
    xmax = 15,
    bar width = 5pt,
    xticklabels=\empty,
    ytick style={draw=none},
    ymajorgrids=true, 
    grid style=dashed, 
	yticklabels={,,},
    xtick style={draw=none},
    label style={font=\Huge}, 
    title style={font=\Huge},
]



\addplot[ 
    color=black, 
    fill=amethyst,
    ybar,
    ] 
    coordinates { 
        (.5, 69)(3, 58)(5.5, 55)(8, 65)(10.5, 61)(13,69)
    }; 

\addplot[ 
    color=black, 
    fill=aureolin,
    ybar,
    ] 
    coordinates { 
        (3.5, 61)(6, 56)(8.5, 58)(11, 56)(13.5,66)
    }; 
   
\addplot[ 
    color=black, 
    fill=persianred,
    ybar,
    ] 
    coordinates { 
        (4, 70)(6.5, 50)(9,62 )(11.5, 57)(14,69)
    };

\addplot[ 
    color=black,  
    fill=parisgreen,
    ybar,
    ] 
    coordinates { 
        (4.5, 78)(7, 50)(9.5, 67)(12, 62)(14.5,74)
    };  
 
\node[anchor=north west] at (rel axis cs:.03,1) {\fontsize{24}{22} \selectfont \textbf{(d)}};

\draw[dashed] (25,0) -- (25,2000);
\draw[dashed] (50,0) -- (50,2000);
\draw[dashed] (75,0) -- (75,2000);
\draw[dashed] (100,0) -- (100,2000);
\draw[dashed] (125,0) -- (125,2000);

\end{axis}
\end{tikzpicture}
}
\end{subfigure}
\begin{subfigure}{.30\textwidth}  
\centering
\label{fusion-edgework:nwr}
\resizebox{\textwidth}{!}{
\begin{tikzpicture}
\tikzstyle{every node}=[font=\Huge]
\begin{axis}[
    style=thick,
    height=7cm, 
    ymin=0, ymax=100, 
    width=14cm,
    xmin = 0,
    xmax = 15,
    bar width = 5pt,
    xticklabels=\empty,
    ytick style={draw=none},
    ymajorgrids=true, 
    grid style=dashed, 
    yticklabels={,,},
    ytick style={draw=none},
    xtick style={draw=none},
    label style={font=\Huge}, 
    title style={font=\Huge},
]



\addplot[ 
    color=black, 
    fill=amethyst,
    ybar,
    ] 
    coordinates { 
        (.5, 73)(3, 57)(5.5, 55)(8, 65)(10.5, 63)(13,66)
    }; 

\addplot[ 
    color=black, 
    fill=aureolin,
    ybar,
    ] 
    coordinates { 
        (3.5, 60)(6, 57)(8.5, 62)(11, 58)(13.5,66)
    };

\addplot[ 
    color=black, 
    fill=persianred,
    ybar,
    ] 
    coordinates { 
        (4, 51)(6.5, 57)(9, 61)(11.5, 58)(14,65)
    };

\addplot[ 
    color=black,  
    fill=parisgreen,
    ybar,
    ] 
    coordinates { 
        (4.5, 62)(7, 71)(9.5, 71)(12, 67)(14.5,67)
    }; 

   
   \node[anchor=north west] at (rel axis cs:.03,1) {\fontsize{24}{22} \selectfont \textbf{(e)}};
    
\draw[dashed] (25,0) -- (25,2000);
\draw[dashed] (50,0) -- (50,2000);
\draw[dashed] (75,0) -- (75,2000);
\draw[dashed] (100,0) -- (100,2000);
\draw[dashed] (125,0) -- (125,2000);

\end{axis}
\end{tikzpicture}
}
\end{subfigure}
\begin{subfigure}{.30\textwidth}  
\centering
\label{fusion-edgework:Radius}
\resizebox{\textwidth}{!}{
\begin{tikzpicture}
\tikzstyle{every node}=[font=\Huge]
\begin{axis}[
    style=thick,
    height=7cm, 
    ymin=0, ymax=100, 
    width=14cm,
    xmin = 0,
    xmax = 15,
    bar width = 5pt,
    xticklabels=\empty,
    ytick style={draw=none},
    ymajorgrids=true, 
    grid style=dashed, 
    yticklabels={,,},
    ytick style={draw=none},
    xtick style={draw=none},
    label style={font=\Huge}, 
    title style={font=\Huge},
]


\addplot[ 
    color=black, 
    fill=amethyst,
    ybar,
    ] 
    coordinates { 
        (.5, 53)(3, 64)(5.5, 55)(8, 69)(10.5, 55)(13,64)
    }; 

\addplot[ 
    color=black, 
    fill=aureolin,
    ybar,
    ] 
    coordinates { 
        (3.5, 61)(6, 60)(8.5, 70)(11, 53)(13.5,59)
    };

\addplot[ 
    color=black, 
    fill=persianred,
    ybar,
    ] 
    coordinates { 
        (4, 60)(6.5, 57)(9, 57)(11.5, 55)(14,65)
    };

\addplot[ 
    color=black,  
    fill=parisgreen,
    ybar,
    ] 
    coordinates { 
        (4.5, 69)(7, 55)(9.5, 74)(12, 62)(14.5,67)
    };
    
    \node[anchor=north west] at (rel axis cs:.03,1) {\fontsize{24}{22} \selectfont \textbf{(f)}};
   
\draw[dashed] (25,0) -- (25,2000);
\draw[dashed] (50,0) -- (50,2000);
\draw[dashed] (75,0) -- (75,2000);
\draw[dashed] (100,0) -- (100,2000);
\draw[dashed] (125,0) -- (125,2000);

\end{axis}
\end{tikzpicture}
}
\end{subfigure}
\caption{Edge-work ratio: normalized \# of edges processed by the fused version over the unfused version in unweighted graphs (a, b, c) and
weighted graphs (d, e, f). (a) and (d) show $\WSP$. (b) and (e) show $\NWR$. (c) and (f) show $\Radius$. 
\label{figure:fusion-edgework-weighted-all}
}
\end{figure*}

\textbf{Fusion Types. \ }
In order to study the performance benefits of the different fusion types that the fusion rules represent,
we compare the unfused and the fused implementations of three representative use-cases \irule{WSP}, \irule{NWR} and \irule{Radius} (presented in \autoref{fig:benchs-spec}).
\autoref{figure:fusion-edgework-weighted-all} shows
the number of edges processed by the
synthesized (fused) programs normalized w.r.t. that by the 
unfused versions,
that we call edge-work ratio.
%
We visit the use-cases and the applied fusion rules (from \autoref{sec:fusion})
in turn.

\textit{\irule{WSP}. \ }
%
\irule{WSP} is fused by the rule \irule{FPNest}.
As described earlier in \autoref{sec:fusion}, this rule fuses
nested path-based reductions. 
The unfused program for this use-case consists of two computation
phases over the edges of the input graph, one after the other.
The first phase calculates the shortest paths from the given source to all the vertices, 
and the second phase computes the capacity of the widest path across the shortest paths.
The fused program, however, executes the two computations in one pass over a pair of values.

\textit{Assessment. \ }
\autoref{figure:fusion-edgework-weighted-all}a (for unweighted graphs)
and
\autoref{figure:fusion-edgework-weighted-all}d (for weighted graphs)
show the edge-work ratio for \irule{WSP}.
%
In unweighted graphs, the fused program processes half the number of edges 
processed by the unfused program.
Similarly, on weighted graphs, the fused program processes 50-70\% of the edges 
processed by the unfused program.
When graphs are unweighted, each edge represents a unit cost 
(that can be either weight or capacity).
In each iteration,
the set of edges that contribute to the weight and capacity values of a vertex are the same.
%
%
Hence, the fused program exploits this overlap 
by simultaneously propagating the two values across each edge;
%
%
%
which reduces passes over the edges by 50\%. 
However,
for weighted graphs, 
the two values
can be propagated to the vertex
in different iterations since the \texttt{min} and \texttt{max}
reductions result in different paths based on different
edge weights.
Hence, 
the processing of 
edges only partially overlap.


\textit{\irule{NWR}. \ }
The unfused version of \irule{NWR}
calculates the narrowest and the widest paths separately.
The two are fused by the rule \irule{FMPair}. 
This rule fuses multiple path-based reductions into a single path-based reduction;
it translates the reduction functions to a single reduction function that operates on pairs. 


\textit{Assessment. \ }
\autoref{figure:fusion-edgework-weighted-all}b (for unweighted graphs)
and
\autoref{figure:fusion-edgework-weighted-all}e (for weighted graphs)
show 
the edge-work ratio for \irule{NWR}.
Similar to \irule{WSP}, the fused program reduces the number of processed edges by 50\% for unweighted graphs and by 51-73\% for weighted graphs.
The fused version propagates the narrowest and widest values over an edge at the same time.
Thus, it benefits from both overlapping propagations and locality of the memory accesses.


\textit{\irule{Radius}. \ }
\irule{Radius} is fused by the rule \irule{FMPair} that we considered above and the rule \irule{FRPair}. 
The rule \irule{FRPair} fuses multiple vertex-based reductions into a single reduction. 
\irule{Radius} computes eccentricity 
(i.e. the maximum shortest distance) by sampling a set of sources. 
We sample two source vertices.
The unfused version computes eccentricity separately for each source.
However, the rules \irule{FMPair} and \irule{FRPair} fuse 
the path-based and vertex-based reductions
across the sources
to a single path-based and a single vertex-based reduction.

\begin{figure*}[t]
\centering
\begin{subfigure}{0.08\textwidth}  
\centering
\vspace{.4cm}
\begin{tikzpicture}
\tikzstyle{every node}=[font=\small]
\begin{axis}[
    style=thin,
    height=3.24cm,  
    width=2cm,
    ymin=0, 
    ymax=100,
    xmin = 0,
    xmax = 1,
    ytick={0, 20, 40, 60, 80, 100},
    ylabel={\fontsize{8}{6} \selectfont Normalized \# \\ of ops (\%)},
    ylabel style={align=center,at={(1,-.82)},xshift=2.15cm},
    ytick style={draw=none},
    xtick style={draw=none},
    x tick label style={draw=none},
    axis line style={draw=none},
    hide x axis,        
    axis y line*=left,
]
\end{axis}
\end{tikzpicture}
\end{subfigure}
\hspace{-.15cm}
\begin{subfigure}{0.30\textwidth}  
\centering
\resizebox{\textwidth}{!}{
\begin{tikzpicture}
\tikzstyle{every node}=[font=\Huge]
\begin{axis}[
	style=thick,
     height=7cm, 
    ymin=0, ymax=100, 
    width=14cm,
    xmin = 0,
    xmax = 15,
    bar width = 5pt,
    xtick={1.25,3.75,6.25,8.75,11.25, 13.75},
    xticklabels={\textbf{Ligra},\textbf{Grid}\\ \textbf{Graph},\textbf{Gemini},\textbf{PG}\\ \textbf{Push},\textbf{PG}\\ \textbf{Pull},\textbf{GraphIt}},
    x tick label style={yshift=7.1cm, align=center},
    ytick style={draw=none},
    ymajorgrids=true, 
    grid style=dashed, 
    yticklabels={,,},
    ytick style={draw=none},
    xtick style={draw=none},
    label style={font=\Huge}, 
    title style={font=\Huge},
]


\addplot[ 
    color=black, 
    fill=amethyst,
    ybar,
    ] 
    coordinates { 
        (.5, 50)(3, 50)(5.5, 50)(8, 50)(10.5, 50)(13,50)
    }; 

\addplot[ 
    color=black, 
    fill=aureolin,
    ybar,
    ] 
    coordinates { 
        (3.5, 50)(6, 50)(8.5, 50)(11, 50)(13.5,50)
    };

\addplot[ 
    color=black, 
    fill=persianred,
    ybar,
    ] 
    coordinates { 
        (4, 50)(6.5, 50)(9, 50)(11.5, 50)(14,50)
    };

\addplot[ 
    color=black,  
    fill=parisgreen,
    ybar,
    ] 
    coordinates { 
        (4.5, 50)(7, 50)(9.5, 50)(12, 50)(14.5,50)
    }; 
 
   \node[anchor=north west] at (rel axis cs:.03,1) {\fontsize{24}{22} \selectfont \textbf{(a)}};

\draw[dashed] (25,0) -- (25,2000);
\draw[dashed] (50,0) -- (50,2000);
\draw[dashed] (75,0) -- (75,2000);
\draw[dashed] (100,0) -- (100,2000);
\draw[dashed] (125,0) -- (125,2000);

\end{axis}
\end{tikzpicture}
}
\end{subfigure}
\begin{subfigure}{0.30\textwidth}  
\centering
\label{fusion-edgework:nwr}
\resizebox{\textwidth}{!}{
\begin{tikzpicture}
\tikzstyle{every node}=[font=\Huge]
\begin{axis}[
    style=thick,
     height=7cm, 
    ymin=0, ymax=100, 
    width=14cm,
    xmin = 0,
    xmax = 15,
    bar width = 5pt,
    xtick={1.25,3.75,6.25,8.75,11.25, 13.75},
    xticklabels={\textbf{Ligra},\textbf{Grid}\\ \textbf{Graph},\textbf{Gemini},\textbf{PG}\\ \textbf{Push},\textbf{PG}\\ \textbf{Pull},\textbf{GraphIt}},
    x tick label style={yshift=7.1cm, align=center},
    ytick style={draw=none},
    ymajorgrids=true, 
    grid style=dashed, 
    yticklabels={,,},
    ytick style={draw=none},
    xtick style={draw=none},
    label style={font=\Huge}, 
    title style={font=\Huge},
]


\addplot[ 
    color=black, 
    fill=amethyst,
    ybar,
    ] 
    coordinates { 
        (.5, 50)(3, 50)(5.5, 50)(8, 50)(10.5, 50)(13,50)
    }; 

\addplot[ 
    color=black, 
    fill=aureolin,
    ybar,
    ] 
    coordinates { 
        (1, 0)(3.5, 50)(6, 50)(8.5, 50)(11, 50)(13.5,50)
    };

\addplot[ 
    color=black, 
    fill=persianred,
    ybar,
    ] 
    coordinates { 
        (1.5, 0)(4, 50)(6.5, 50)(9, 50)(11.5, 50)(14,50)
    };

\addplot[ 
    color=black,  
    fill=parisgreen,
    ybar,
    ] 
    coordinates { 
        (2, 0)(4.5, 50)(7, 50)(9.5, 50)(12, 50)(14.5,50)
    }; 


   \node[anchor=north west] at (rel axis cs:.03,1) {\fontsize{24}{22} \selectfont \textbf{(b)}};
    
\draw[dashed] (25,0) -- (25,2000);
\draw[dashed] (50,0) -- (50,2000);
\draw[dashed] (75,0) -- (75,2000);
\draw[dashed] (100,0) -- (100,2000);
\draw[dashed] (125,0) -- (125,2000);

\end{axis}
\end{tikzpicture}
}
\end{subfigure}
\begin{subfigure}{0.30\textwidth}  
\centering
\label{fusion-edgework:Radius}
\resizebox{\textwidth}{!}{
\begin{tikzpicture}
\tikzstyle{every node}=[font=\Huge]
\begin{axis}[
    style=thick,
     height=7cm, 
    ymin=0, ymax=100, 
    width=14cm,
    xmin = 0,
    xmax = 15,
    bar width = 5pt,
    xtick={1.25,3.75,6.25,8.75,11.25, 13.75},
    xticklabels={\textbf{Ligra},\textbf{Grid}\\ \textbf{Graph},\textbf{Gemini},\textbf{PG}\\ \textbf{Push},\textbf{PG}\\ \textbf{Pull},\textbf{GraphIt}},
    x tick label style={yshift=7.1cm, align=center},
    ytick style={draw=none},
    ymajorgrids=true, 
    grid style=dashed, 
    yticklabels={,,},
    ytick style={draw=none},
    xtick style={draw=none},
    label style={font=\Huge}, 
    title style={font=\Huge},
]


\addplot[ 
    color=black, 
    fill=amethyst,
    ybar,
    ] 
    coordinates { 
        (.5, 57)(3, 57)(5.5, 66)(8, 72)(10.5, 54)(13,70)
    }; 

\addplot[ 
    color=black, 
    fill=aureolin,
    ybar,
    ] 
    coordinates { 
        (3.5, 52)(6, 74)(8.5, 72)(11, 54)(13.5,60)
    };

\addplot[ 
    color=black, 
    fill=persianred,
    ybar,
    ] 
    coordinates { 
        (4, 54)(6.5, 74)(9, 79)(11.5, 61)(14,60)
    };

\addplot[ 
    color=black,  
    fill=parisgreen,
    ybar,
    ] 
    coordinates { 
        (4.5, 53)(7, 67)(9.5, 68)(12, 72)(14.5,61)
    }; 
    
    \node[anchor=north west] at (rel axis cs:.03,1) {\fontsize{24}{22} \selectfont \textbf{(c)}};
   
\draw[dashed] (25,0) -- (25,2000);
\draw[dashed] (50,0) -- (50,2000);
\draw[dashed] (75,0) -- (75,2000);
\draw[dashed] (100,0) -- (100,2000);
\draw[dashed] (125,0) -- (125,2000);

\end{axis}
\end{tikzpicture}
}
\end{subfigure}
\end{figure*}

\textit{Assessment. \ }
\autoref{figure:fusion-edgework-weighted-all}c (for unweighted graphs)
and
\autoref{figure:fusion-edgework-weighted-all}f (for weighted graphs)
show the edge-work ratio for \irule{Radius}.
%
We observe that on unweighted graphs, 
the fused version processes 52-78\% of the number of edges
that the unfused version processes.
This ratio is 53-74\% on weighted graphs. 
Even though fusion enables computation of multiple eccentricity values at the same time, contrary to \irule{WSP} and \irule{NWR},
we do not observe the 50\% reduction.
This is because eccentricity computations across different sources can occur via paths that don't necessarily overlap. 
The fused version exploits the partial overlaps.
%

%
%
We observe that the reduction in edge computations is different across different frameworks as well. 
For example, 
the edge-work ratio
is
52-68\%
in GridGraph,
whereas
54-78\%
in PowerGraph.
This is because of the difference in the scheduling strategies across these different frameworks, that lead to different overlaps in edge computations. 
This means that even if the same edge propagates values for multiple sources, 
certain frameworks may schedule processing of that edge for different sources in different iterations.

Next, we consider the edge-work ratio and absolute execution times for more elaborate use-cases with multiple fusions.
(Due to space limitation, the absolute execution times 
for the simpler use-cases above,
\irule{WSP}, \irule{NWR} and \irule{Radius},
are available in \appref{\secEvaluationFusionExec}.
Further,
more experiments 
including
the scalability of the fusion process on increasing number of sources in the \irule{Radius} use-case
are available in 
\appref{\secEvaluationFusionScale}.)




\begin{table}
\renewcommand\thetable{3} 
\centering
\scriptsize
\caption{\label{table:more-exec-time}Execution times (in seconds). H: Handwritten, S: Synthesized, R: the ratio $\frac{H}{S}$.
}
\resizebox{\textwidth}{!}{
\begin{tabular}{c | c | c | c | c | c | c | c | c | c | c | c | c | c | c | c | c | c | c | c }
\toprule
\centering
\multirow{2}{*}{Prog.} & \multirow{2}{*}{Input} & \multicolumn{3}{c|}{Ligra} & \multicolumn{3}{c|}{GridGraph}& \multicolumn{3}{c|}{Gemini} & \multicolumn{3}{c|}{PowerGraph (Push)} & \multicolumn{3}{c|}{PowerGraph (Pull)} & \multicolumn{3}{c}{GraphIt (Push)}\\
\cline{3-20}\\[-2mm]
 & & H & S & \textbf{R} & H & S & \textbf{R} & H & S & \textbf{R} & H & S & \textbf{R} & H & S & \textbf{R} & H & S & \textbf{R}\\
 \cline{1-20}\\[-2mm]
$\multirow{4}{*}{DRR}$  & LJ & 13.1 & 4 & \textbf{3.2} & 15.3 & 3.8 & \textbf{4} & 0.9 & 0.3 & \textbf{3} &  20.4 & 6.4 & \textbf{3.2} &  36 & 10 & \textbf{3.6}& 0.35 &1 & \textbf{2.8}\\
  & TW  & - & - & - & 82 & 23 & \textbf{3.6} & 11.1 & 5.5 & \textbf{2} &  120 & 48 & \textbf{2.5} & 292 & 81 & \textbf{3.6}& 6.7 & 20.7& \textbf{3}\\
  & TM  & - & - & - & 141 & 44 & \textbf{3.3} & 18.2 & 6.3 & \textbf{2.9} & 166 & 50 & \textbf{3.3} & 462 & 86 & \textbf{2.9}& 12.2 & 34.7 & \textbf{3.5}\\
  & FR  & - & - & - & 265 & 73 & \textbf{3.6} & 36 & 17 & \textbf{2} & 247 & 86 & \textbf{2.9} & 522 & 154 & \textbf{3.4}& 17 & 47 & \textbf{3.5}\\
 \bottomrule\\[-2mm]
$\multirow{4}{*}{Trust}$  & LJ & 12.3 & 6 & \textbf{2.05} & 14.1 & 4.4 &  \textbf{3.2} & 0.76 & 0.37 & \textbf{2} & 20.5 & 7.5& \textbf{2.7} & 37 & 10 &  \textbf{3.7}& 0.47 & 1& \textbf{2.2}\\
  & TW & - & - & - & 85 & 28 & \textbf{3.1} & 10.5 & 6.6 & \textbf{1.6} &  122 & 50 & \textbf{2.4} & 293 & 99 & \textbf{2.9} & 11.7 &22 & \textbf{2}\\
  & TM & - & - & - & 122 & 40 & \textbf{3} & 14.5 & 8 & \textbf{1.8} & 157 & 71 & \textbf{2.2} & 455 & 129 & \textbf{3.5} & 23.3 &80 & \textbf{3.4}\\
  & FR & - & - & - & 218 & 117& \textbf{2} & 34.5 & 16 & \textbf{2.2} & 252 & 98 & \textbf{2.6} & 526 & 173 & \textbf{3}& 25 & 52 & \textbf{2}\\
\bottomrule\\[-2mm]
$\multirow{4}{*}{RDS}$  & LJ & 14.4 & 7.8 & \textbf{1.8} & 11.5 & 6 & \textbf{1.9} & 0.9 & 0.6 & \textbf{1.5} & 23 & 15.1 & \textbf{1.5} & 41 & 28 & \textbf{1.4}&  0.73& 1& \textbf{1.4}\\
 & TW  & - & - & - & 74 & 47 & \textbf{1.6} & 11 & 6 & \textbf{1.8} & 130 & 108 & \textbf{1.2} & - & - & \textbf{-}& 13.2 & 20.5& \textbf{1.5}\\
 & TM & - & - & - & 142 & 82 & \textbf{1.7} & 11 &  7.8 & \textbf{1.4} & 200 & 134 & \textbf{1.5} & - & - & \textbf{-}& 23 & 45.7& \textbf{2}\\
 & FR & - & - & - & 210 & 175 & \textbf{1.2} & 37 & 19 & \textbf{1.9} & 286 & 198 & \textbf{1.5} & - & - & \textbf{-}& 19 & 35 & \textbf{1.7}\\
\bottomrule
\end{tabular}
}
\end{table}
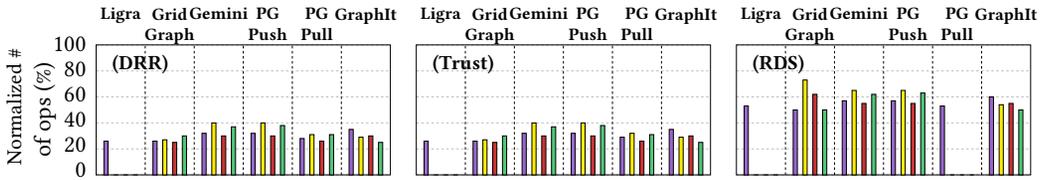
\begin{figure*}[t]
\vspace{-.3cm}
\begin{subfigure}{0.08\textwidth}  
\centering
\vspace{.4cm}
\begin{tikzpicture}
\tikzstyle{every node}=[font=\small]
\begin{axis}[
    style=thin,
    height=3.24cm,  
    width=2cm,
    ymin=0, 
    ymax=100,
    xmin = 0,
    xmax = 1,
    ytick={0, 20, 40, 60, 80, 100},
    ylabel={\fontsize{8}{6} \selectfont Normalized \# \\ of ops (\%)},
    ylabel style={align=center,at={(.9,-.82)},xshift=2.15cm},
    ytick style={draw=none},
    xtick style={draw=none},
    x tick label style={draw=none},
    axis line style={draw=none},
    hide x axis,        
    axis y line*=left,
]
\end{axis}
\end{tikzpicture}
\end{subfigure}
\centering
\begin{subfigure}{.30\textwidth}  
\centering
\resizebox{\textwidth}{!}{
\begin{tikzpicture}
\tikzstyle{every node}=[font=\Huge]
\begin{axis}[
	style=thick,
     height=7cm, 
    ymin=0, ymax=100, 
    width=14cm,
    xmin = 0,
    xmax = 15,
    bar width = 5pt,
    xtick={1.25,3.75,6.25,8.75,11.25, 13.75},
    xticklabels={\textbf{Ligra},\textbf{Grid}\\ \textbf{Graph},\textbf{Gemini},\textbf{PG}\\ \textbf{Push},\textbf{PG}\\ \textbf{Pull},\textbf{GraphIt}},
    x tick label style={yshift=7.1cm, align=center},
    ytick style={draw=none},
    ymajorgrids=true, 
    grid style=dashed, 
    yticklabels={,,},
    ytick style={draw=none},
    xtick style={draw=none},
    label style={font=\Huge}, 
    title style={font=\Huge},
]

\addplot[ 
    color=black, 
    fill=amethyst,
    ybar,
    ] 
    coordinates { 
        (.5, 26)(3, 26)(5.5, 32)(8, 32)(10.5, 28)(13,35)
    }; 

\addplot[ 
    color=black, 
    fill=aureolin,
    ybar,
    ] 
    coordinates { 
        (1, 0)(3.5, 27)(6, 40)(8.5, 40)(11, 31)(13.5,29)
    };

\addplot[ 
    color=black, 
    fill=persianred,
    ybar,
    ] 
    coordinates { 
		(1.5, 0)(4, 25)(6.5, 30)(9, 30)(11.5, 26)(14,30)
    };

\addplot[ 
    color=black,  
    fill=parisgreen,
    ybar,
    ] 
    coordinates { 
        (2, 0)(4.5, 30)(7, 37)(9.5, 38)(12, 31)(14.5,25)
    }; 
 
   \node[anchor=north west] at (rel axis cs:.03,.95) {\fontsize{24}{22} \selectfont \textbf{(DRR)}};

\draw[dashed] (25,0) -- (25,2000);
\draw[dashed] (50,0) -- (50,2000);
\draw[dashed] (75,0) -- (75,2000);
\draw[dashed] (100,0) -- (100,2000);
\draw[dashed] (125,0) -- (125,2000);

\end{axis}
\end{tikzpicture}
}
\end{subfigure}
\begin{subfigure}{.30\textwidth}  
\centering
\label{fusion-edgework:nwr}
\resizebox{\textwidth}{!}{
\begin{tikzpicture}
\tikzstyle{every node}=[font=\Huge]
\begin{axis}[
    style=thick,
     height=7cm, 
    ymin=0, ymax=100, 
    width=14cm,
    xmin = 0,
    xmax = 15,
    bar width = 5pt,
    xtick={1.25,3.75,6.25,8.75,11.25, 13.75},
    xticklabels={\textbf{Ligra},\textbf{Grid}\\ \textbf{Graph},\textbf{Gemini},\textbf{PG}\\ \textbf{Push},\textbf{PG}\\ \textbf{Pull},\textbf{GraphIt}},
    x tick label style={yshift=7.1cm, align=center},
    ytick style={draw=none},
    ymajorgrids=true, 
    grid style=dashed, 
    yticklabels={,,},
    ytick style={draw=none},
    xtick style={draw=none},
    label style={font=\Huge}, 
    title style={font=\Huge},
]


\addplot[ 
    color=black, 
    fill=amethyst,
    ybar,
    ]  
    coordinates { 
        (.5, 26)(3, 26)(5.5, 32)(8, 32)(10.5, 29)(13,35)
    }; 

\addplot[ 
    color=black, 
    fill=aureolin,
    ybar,
    ] 
    coordinates { 
        (1, 0)(3.5, 27)(6, 40)(8.5, 40)(11, 32)(13.5,29)
    };

\addplot[ 
    color=black, 
    fill=persianred,
    ybar,
    ] 
    coordinates { 
        (1.5, 0)(4, 25)(6.5, 30)(9, 30)(11.5, 26)(14,30)
    };

\addplot[ 
    color=black,  
    fill=parisgreen,
    ybar,
    ] 
    coordinates { 
        (2, 0)(4.5, 30)(7, 37)(9.5, 38)(12,31)(14.5,25)
    }; 


   \node[anchor=north west] at (rel axis cs:.03,.95) {\fontsize{24}{22} \selectfont \textbf{(Trust)}};
    
\draw[dashed] (25,0) -- (25,2000);
\draw[dashed] (50,0) -- (50,2000);
\draw[dashed] (75,0) -- (75,2000);
\draw[dashed] (100,0) -- (100,2000);
\draw[dashed] (125,0) -- (125,2000);

\end{axis}
\end{tikzpicture}
}
\end{subfigure}
\begin{subfigure}{.30\textwidth}  
\centering
\label{fusion-edgework:Radius}
\resizebox{\textwidth}{!}{
\begin{tikzpicture}
\tikzstyle{every node}=[font=\Huge]
\begin{axis}[
    style=thick,
     height=7cm, 
    ymin=0, ymax=100, 
    width=14cm,
    xmin = 0,
    xmax = 15,
    bar width = 5pt,
    xtick={1.25,3.75,6.25,8.75,11.25, 13.75},
    xticklabels={\textbf{Ligra},\textbf{Grid}\\ \textbf{Graph},\textbf{Gemini},\textbf{PG}\\ \textbf{Push},\textbf{PG}\\ \textbf{Pull},\textbf{GraphIt}},
    x tick label style={yshift=7.1cm, align=center},
    ytick style={draw=none},
    ymajorgrids=true, 
    grid style=dashed, 
    yticklabels={,,},
    ytick style={draw=none},
    xtick style={draw=none},
    label style={font=\Huge}, 
    title style={font=\Huge},
]


\addplot[ 
    color=black, 
    fill=amethyst,
    ybar,
    ] 
    coordinates { 
        (.5, 53)(3, 50)(5.5, 57)(8, 57)(10.5, 53)(13,60)
    }; 

\addplot[ 
    color=black, 
    fill=aureolin,
    ybar,
    ] 
    coordinates { 
        (1, 0)(3.5, 73)(6, 65)(8.5, 65)(11, 0)(13.5,54)
    };

\addplot[ 
    color=black, 
    fill=persianred,
    ybar,
    ] 
    coordinates { 
        (1.5, 0)(4, 62)(6.5, 55)(9, 55)(11.5, 0)(14,55)
    };

\addplot[ 
    color=black,  
    fill=parisgreen,
    ybar,
    ] 
    coordinates { 
        (2, 0)(4.5, 50)(7, 62)(9.5, 63)(12, 0)(14.5,50)
    }; 
    
    \node[anchor=north west] at (rel axis cs:.03,.95) {\fontsize{24}{22} \selectfont \textbf{(RDS)}};
   
\draw[dashed] (25,0) -- (25,2000);
\draw[dashed] (50,0) -- (50,2000);
\draw[dashed] (75,0) -- (75,2000);
\draw[dashed] (100,0) -- (100,2000);
\draw[dashed] (125,0) -- (125,2000);

\end{axis}
\end{tikzpicture}
}
\end{subfigure}

\caption{Edge-work ratio: Normalized \# of edges processed by the fused version over the unfused version.\label{figure:more-fusion-edgework-separate-all}
}
\end{figure*}

%
%
%
%
\textbf{Multiple Fusions. \ }
%
We study the performance benefits of fusion on
more elaborated 
use-cases: \irule{Trust}, \irule{DRR}, and \irule{RDS} (presented in \autoref{fig:benchs-spec}).
We report both 
edge-work ratio
and absolute execution times in \autoref{figure:more-fusion-edgework-separate-all} and \autoref{table:more-exec-time} respectively.   
Our experimental results show that fusion reduces
the edge-work ratio to a quarter
and leads to up-to
4$\times$ speedup.

\textit{\irule{DRR}. \ }
\irule{DRR} calculates the ratio of the diameter over radius 
sampled over two sources. 
In addition to the rules \irule{FMPair}, \irule{FRPair} and \irule{FLetsBin} which fuse path-based and vertex-based reductions, common operation elimination rules (\appref{\secComSubexpElim}) 
eliminate redundant path-based computations in diameter and radius.
Therefore, instead of 4 reductions, \name{} fuses and calculates 1 reduction
(\apprefp{\secExampleFusions}).
In \autoref{figure:more-fusion-edgework-separate-all}, we observe that
the edge-work ratio is 25-40\%.
This translates to 2-$4\times$ speedup in \autoref{table:more-exec-time}.
Note that the theoretical bound on the edge-work ratio
is 25\%, which is achieved when the path-based computations 
for the two 
sources fully overlap.

\textit{\irule{Trust}. \ }
\irule{Trust}
specifies
the trust from a given set of nodes to other nodes.
It 
applies
division and maximum operators between path-based reductions: the widest and shortest paths.
The rules \irule{FILetBin} and \irule{FMPair} fuse the 4 path-based reductions to 1.
As shown in \autoref{figure:more-fusion-edgework-separate-all}, 
the edge-work ratio is 25-40\%,
and
\autoref{table:more-exec-time} shows,
the speedup across different frameworks is 1.6-$3.7\times$.
The theoretical bound on 
the edge-work ratio
is again
25\%, similar to the \irule{DRR} use-case.

\textit{\irule{RDS}. \ }
Given a source $s$,
\irule{RDS} calculates
the narrowest of the
widest paths
to vertices
within the radius neighbourhood of $s$.
\irule{RDS} has a nested reduction for \irule{Radius}
and
is transformed
by fusion rules for nested vertex-based reductions
(\apprefp{\secMultipleRounds}).
The inner \irule{Radius} is factored and fused as before.
Moreover, the two path-based reductions, the narrowest and shortest paths, are fused by the rules \irule{FILetBin} and \irule{FMPair}. 
This results in a sequence of two iteration-map-reduce rounds.
The theoretical bound for
the edge-work ratio
is 50\% mainly because the fused and unfused programs perform two and four sequences of iteration-map-reduce rounds respectively. 
\autoref{figure:more-fusion-edgework-separate-all} shows that 
the edge-work ratio is 57-85\%
which translates to 1.2-$2\times$ speedup 
in \autoref{table:more-exec-time}.

\begin{wrapfigure}[16]{R}{0.31\textwidth}
\small
\centering
\begin{threeparttable}
 \begin{tabular}{|p{1.1cm}|p{.8cm}|p{.4cm}|p{.5cm}|} 
 \hline
 Program & \#PBR\tnote{1} & F\tnote{2} & CS\tnote{3} \\ [0.5ex] 
 \hline\\[-2.5mm]
 \irule{BFS} & 1 & 51 & 25\\ 
 \irule{CC} & 1 & 3 & 1\\
 \irule{SSSP} & 1 & 2 & 24\\
 $\WP$ & 1 & 2 & 29 \\ 
 $\textsf{PR}$ & - & - & - \\ 
 \irule{WSP} & 2 & 93 & 44 \\ 
 \irule{NWR} & 2 & 2 & 58 \\ 
 \irule{Radius} & 2 & 2 & 49 \\ 
 \irule{DS} & 2 & 2 & 49 \\ 
  \irule{DRR} & 4 & 3 & 50 \\ 
   \irule{Trust} & 4 & 3 & 105 \\ 
 \irule{RDS} & 4 & 10 & 102\\

 \hline
 \end{tabular}
 \begin{tablenotes}
\item[1] \# of path-based reductions
\item[2] Fusion time (ms)
\item[3] Constraint solving time (s)
\end{tablenotes}
\end{threeparttable}
\caption{\label{table:synth-time}Analysis Time}
\end{wrapfigure}
\textbf{Synthesis time. \ }
%
\autoref{table:synth-time} presents the synthesis time
that is
below 2 minutes and often seconds.
The use cases that require desugaring (\irule{BFS}) or fusion of nested reductions (\irule{WSP}) take more time.

\mclearpage
\section{Related Work}

\textbf{Graph Processing Frameworks and Synthesis. \ }
Graph processing systems provide 
interfaces
to hide the implementation details such as parallelism, synchronization and communication in scalable runtimes. At the heart of graph computations are operations over vertex and edge values and scheduling policies to determine the order in which operations are performed. Parallelism is often extracted at the vertex and edge level, and hence, most interfaces
allow computations to be directly expressed as vertex-level and edge-level operations~\cite{pregel,graphlab,distributedgraphlab,powergraph,ligra,gemini,grazelle,gridgraph,xstream,graphit,galois,hoang2019round,gluon}. 
Certain DSLs raise the level of abstraction in order to simplify development of graph algorithms~\cite{greenmarl,emptyheaded,gremlin,pgql,pgqlgreenmarl}. 
%
%
%
%
Contrary to our synthesis process,
graph processing DSLs
\cite{dhfalcon,lighthouse,Abelian,fregel}
require users to 
write vertex- or edge-level kernel functions. 
However, they provide implementations 
for different architectures such as GPUs and distributed platforms.
Further, they generate implementations that are tied to their runtime specifics.
In the synthesis domain, Elixir \cite{elixir,automatedplanning} synthesizes 
multiple 
parallel 
implementations from 
the specification of
a graph computation and 
applies automated reasoning to optimize them.
In contrast,
$\name$ offers 
a more high-level
specification language
and
automatically synthesizes the kernel functions.

\textbf{Program Synthesis. \ }
Program synthesis has always been an area of interest for computer scientists.
%
Previous works have employed enumeration 
\cite{transit,itzhaky2010simple},
variants of syntax-guided synthesis \cite{sygus}
%
and type-guided synthesis \cite{osera2015type,polikarpova2016program}
to synthesize
protocol snippets \cite{transit}
and
Excel macros \cite{gulwani2011automating, gulwani2012spreadsheet}. 
$\name$'s 
synthesis process enumerates graph processing kernel functions
based on a syntax grammar for local computations.
Previous works have also used
constraint solving
to fill holes in program sketches 
\cite{sketch, bitsketch}
including 
architectural kernel functions
\cite{msl},
and
to 
synthesize
control structures,
imperative programs~\cite{srivastava2010program,feng2017} and program templates~\cite{barman2015toward}
 or to compose APIs 
\cite{jha2010oracle,frangel}.
The $\name$ synthesis tool applies SMT solvers to check that
the candidate kernel functions
satisfy the correctness conditions
of the iterative models.
Built on top of Fregel, \cite{akimasa2018} uses SAT solvers to
optimize 
kernel functions.
In contrast, $\name$ automatically synthesize the kernel functions.
%
%
%
%
Superoptimization is another thread of synthesis 
which applies stochastic search methods
to synthesize programs \cite{superoptimizer,joshi2002denali,joshi2006denali,bansal2006automatic,schkufza2013stochastic}. 
Moreover, Souper~\cite{sasnauskas2017souper} took a step further by synthesizing superoptimizers.
%
In contrary to superoptimization
which focuses on optimizing machine-level code, 
$\name$ fusion rules optimize high-level graph processing specifications.
Program synthesis has been also utilized
to synthesize distributed programs \cite{mapreduce,transit,hamsaz}.

\textbf{Fusion. \ }
Fusion is a versatile optimization technique.
Loop fusion~\cite{darte1999complexity,kennedy1993maximizing,qasem2006,bondhugula2008}
merges the bodies of loops on regular structures such as arrays
and hence
reduces the number of memory accesses and improves locality.
Fusion also has been applied to tree structures
\cite{rajbhandari2016fusing,rajbhandari2016domain,sakka2017treefuser,sakka2019sound}
to combine multiple phases of traversal
or fuse different stages of
data processing pipelines~\cite{Saarikivi17}
to enhance data locality.
%
%
%
%
%
Deforestation of functional programs \cite{wadler1988,gill1993,chin1992safe}
combines a sequence of function applications into a single function application
and eliminates intermediate values.
However, deforestation is oblivious to the primitives of graph computation.
%
Graph computations use three fundamental primitives; thus,
we structure these primitives as the triple-let term.
The fusion rules 
transform the computations to this structure
and maintain it during fusion.

\section{Conclusion}
\label{sec:conclusion}

We saw \name, a graph analytics
language and synthesizer.
It features semantics-preserving fusion optimizations.
It automatically synthesizes 
kernel functions
based on 
correctness conditions 
for iterative reductions.
It 
generates 
code for 
high-performance graph processing frameworks.
We hope that it 
motivates further research
to simplify and accelerate data analytics.

%




\end{document}